\documentclass[aps,pre,twocolumn,groupedaddress,showpacs]{revtex4-2}
\newcommand{\bec}[1]{\mbox{\boldmath $ #1$}}
\usepackage{bm}
\usepackage{graphicx}
\usepackage{color}

{}

{}
{}

\newcommand{\meanN}{\overline{n}}
\newcommand{\meann}{\overline{n}}
\newcommand{\meanP}{\overline{P}}
\newcommand{\meanT}{\overline{T}}

\begin{document}
\title{Experimental study of turbulent thermal diffusion of particles
in an inhomogeneous forced convective turbulence}
\author{E.~Elmakies}
\author{O.~Shildkrot}
\author{N. Kleeorin}
\author{A. Levy}
\author{I.~Rogachevskii}
\email{gary@bgu.ac.il}

\bigskip
\affiliation{
The Pearlstone Center for Aeronautical Engineering
Studies, Department of Mechanical Engineering,
Ben-Gurion University of the Negev, P.O.Box 653,
Beer-Sheva 8410530,  Israel}

\date{\today}
\begin{abstract}
We investigate experimentally phenomenon of turbulent thermal diffusion of
micron-size solid particles in an inhomogeneous convective turbulence
forced by one vertically-oriented oscillating grid in an air flow.
This effect causes formation of large-scale inhomogeneities in particle spatial distributions
in a temperature-stratified turbulence.
We perform detailed comparisons of the experimental results
with those obtained in our previous experiments with an inhomogeneous and anisotropic
stably stratified turbulence produced by a one oscillating grid in the air flow.
Since the buoyancy increases the turbulent kinetic energy for convective turbulence
and decreases it for stably stratified turbulence,
the measured turbulent velocities for convective turbulence are larger than those
for stably stratified turbulence.
This tendency is also seen in the measured vertical integral turbulent length scales.
Measurements of temperature and particle number density spatial distributions
show that particles are accumulated in the vicinity
of the minimum of the mean temperature
due to phenomenon of turbulent thermal diffusion.
This effect is observed in both, convective and stably stratified turbulence,
where we find the effective turbulent thermal diffusion coefficient
for micron-size particles.
The obtained experimental results are in agreement with theoretical predictions.
\end{abstract}

\maketitle

\section{Introduction}
\label{sect1}

Turbulent transport of particles has been a subject of many studies
due to numerous applications in geophysics and environmental sciences,
astrophysics, and various industrial flows
\cite{CSA80,ZRS90, BLA97, SP06, ZA08,CST11,RI21}.
Different mechanisms of large-scale and small-scale clustering
of inertial particles have been proposed.
The large-scale clustering occurs in scales which are much
larger than the integral scale of turbulence,
while the small-scale clustering is observed
in scales which are much smaller than the integral turbulence scale.

The large-scale clustering of inertial particles in isothermal non-stratified
inhomogeneous turbulence occurs due to turbophoresis
\cite{CTT75,RE83,G97,EKR98,G08,MHR18},
which is a combined effect of particle inertia and inhomogeneity of turbulence.
Turbophoresis results in appearance of the additional non-diffusive
turbulent flux of inertial particles ${\bm J}_{\rm turboph} = \overline{n}
\, \, \overline{\bm V}_{\rm turboph}$, where the mean particle velocity
caused by turbophoresis can be written as
\begin{eqnarray}
\overline{\bm V}_{\rm turboph} = - \kappa_{\rm turb} \, \, {\bec{\nabla} \left\langle{\bm u}^2\right\rangle \over 2} .
\label{RT10}
\end{eqnarray}
Here $\overline{n}$ is the mean number density of inertial particles,
${\bm u}$ is the turbulent fluid velocity, $\kappa_{\rm turb}$ is the turbophoretic
coefficient which generally depends on the Stokes number
${\rm St} = \tau_p/\tau_\nu$ and the fluid Reynolds number ${\rm Re}= \ell_0\, u_{\rm rms}/\nu$,
where $\tau_\nu=\tau_0/{\rm Re}^{1/2}$ is the Kolmogorov viscous time, $\tau_0=\ell_0 / u_{\rm rms}$
in the characteristic turbulent time, $\nu$ is the kinematic viscosity,
$u_{\rm rms} \equiv \sqrt{\langle{\bm u}^2\rangle}$ is the rms velocity in the integral turbulence scale $\ell_0$
and $\tau_p$ is the Stokes time for the small spherical particles.
Due to turbophoresis, inertial particles are accumulated in the vicinity of the minimum of the turbulent intensity.
In particular, direct numerical simulations (DNS) \cite{MHR18}
show that inertial particles in inhomogeneously
forced isothermal turbulent flows are accumulated at the
minima of turbulent velocity. Two turbulent
transport processes, turbophoresis and turbulent diffusion
determine the spatial distribution of the particles.
Numerical simulations \cite{MHR18} demonstrate  that the non-dimensional product of the turbophoretic coefficient
$\kappa_{\rm turb}$ and the rms velocity $u_{\rm rms}$ increases linearly with the parameter
${\rm St}_f$ for ${\rm St}_f\ll 1$, reaches a maxima for ${\rm St}_f \sim 10$ and decreases
as ${\rm St}_f^{-1/3}$ for large ${\rm St}_f$,
where ${\rm St}_f= \tau_p/\tau_0$ is the Stokes number defined using the characteristic
flow time scale $\tau_0$ based on the forcing scale of turbulence.
The same large-scale clustering phenomenon caused by turbophoresis has been studied in DNS of turbulent Kolmogorov
flows \cite{LCB16}. Although the authors do not interpret their results as a balance
between turbophoretic and turbulent diffusive fluxes, they do observe that the large-scale clustering
increases for small ${\rm St}_f$ but this trend reverses smoothly at
higher values of ${\rm St}_f$.
The large-scale clustering due to turbophoresis has been also observed in DNS
of turbulent channel flows \cite{SSB12} and in various experimental studies
\cite{KHB95,RR04}.

Another example of the large-scale clustering
of inertial particles is the phenomenon of turbulent thermal diffusion
that is a combined effect of the temperature stratified turbulence
and inertia of small particles \cite{EKR96,EKR97}.
Turbulent thermal diffusion is a
purely collective phenomenon occurring in
temperature stratified turbulence and resulting
in the appearance of a non-zero mean effective
velocity of particles in the direction opposite
to the mean temperature gradient.
This implies that this phenomenon causes a non-diffusive
turbulent flux of particles in the direction of the turbulent heat flux.
A competition between the turbulent thermal diffusion and
turbulent diffusion determines the conditions for
the formation of large-scale particle
concentrations in the vicinity of the mean
temperature minimum.

Turbulent thermal diffusion has been intensively investigated analytically
\cite{EKR96,EKR97,EKR00,EKRS00,EKRS01,PM02,RE05,AEKR17}
using different theoretical approaches.
This effect has been detected in DNS \cite{HKRB12,RKB18}.
Turbulent thermal diffusion has been observed in geophysical turbulence, e.g.,  in the atmosphere
of the Earth \cite{SSEKR09} and the atmosphere of Titan \cite{EKPR97},
and it also has been discussed in astrophysical turbulence applications \cite{H16}.
Moreover, the phenomenon of turbulent thermal diffusion has been detected
in laboratory experiments in nearly isotropic and homogeneous turbulence
produced by two oscillating grids
\cite{BEE04,EEKR04,EEKR06a,AEKR17}
and in a multi-fan produced turbulence \cite{EEKR06b}.
Recently the phenomenon of turbulent thermal diffusion
has been found in an inhomogeneous and anisotropic
stably stratified turbulence produced by one oscillating grid in the air flow \cite{EKRL22}.
These experiments have demonstrated formation of inhomogeneous distributions
of micron-size particles in the vicinity of the mean temperature minimum.

The main goal of the present study is to investigate experimentally
the phenomenon of turbulent thermal diffusion of
the micron-size solid particles in an inhomogeneous convective turbulence
forced by one oscillating grid in the air flow.
In the experiments, we measure velocity fields applying Particle Image Velocimetry (PIV).
We measure temperature field with a temperature probe equipped
with 12 E thermocouples.
In addition, we determine spatial distributions of small solid particles
by a PIV system using the effect of the Mie light scattering by particles in the flow.
We perform detailed comparisons of the obtained experimental results
with those in the experiments in an inhomogeneous and anisotropic
stably stratified turbulence produced by one oscillating grid  \cite{EKRL22}
and in a convective turbulence  forced by two oscillating grids in the air flow \cite{EEKR11}.
This paper is organized as follows.
In Sec.~\ref{sect2} we elucidate the mechanism of the phenomenon of turbulent thermal diffusion
and determine the turbulent flux of particles using the spectral $\tau$ approach
for fully developed temperature-stratified turbulence.
In Sec. ~\ref{sect3} we discuss our experimental facilities and instrumentation, and
in Sec. ~\ref{sect4} we describe the obtained experimental results.
Finally,  in Sec.~\ref{sect5} we outline conclusions.

\section{Turbulent thermal diffusion}
\label{sect2}

In this section we determine the turbulent flux of particles in a temperature-stratified 
turbulence and elucidate the mechanism related to the effect of turbulent thermal diffusion.
We study dynamics of small non-inertial particles advected by a turbulent fluid flow.
An evolution of the particle number density $n(t,{\bm x})$
in a fluid velocity field ${\bm U}(t,{\bm x})$ is determined by the convective-diffusion equation:
\begin{eqnarray}
{\partial n \over \partial t} + \bec{\nabla} {\bf \cdot} \left(n \, {\bm U} - D
\bec{\nabla} n \right) = 0 ,
\label{D2}
\end{eqnarray}
where $D= k_B \,T/(6\pi \rho \, \nu \, a_p)$ is the coefficient
of the molecular (Brownian) diffusion of particles having the radius $a_p$.
Here $T$ and $\rho$  are the fluid temperature and
density, respectively and $k_B$ is the Boltzmann constant.
The fluid velocity is a turbulent field produced by, e.g., an external steering force.
Equation~(\ref{D2}) is a conservation law for the total number of particles
that implies that the total number of particles  is conserved in a closed volume.
Here we do not consider a coagulation of particles or chemical reactions
as well as condensation or evaporation of droplets which change the
total number of particles or droplets in a closed volume.

Assuming for simplicity, that the diffusion coefficient is independent of coordinates,
Eq.~(\ref{D2}) can be rewritten as
\begin{eqnarray}
{\partial n \over \partial t} + \bec{\nabla} {\bf \cdot}(n \, {\bm U}) = D \,
\Delta n .
\label{D1}
\end{eqnarray}
We use a point-particle approximation that implies that the size of particles is
very small in comparison with all possible scales of fluid motions.
When the fluid velocity is much less than the sound speed
(i.e., for low-Mach-number fluid flows),
the continuity equation for the fluid density can be used in an anelastic approximation,
$\bec{\nabla} {\bf \cdot}(\rho \, {\bm U}) = 0$.
This equation can be rewritten as
$\bec{\nabla} {\bf \cdot} {\bm U} = - ({\bm U} {\bf \cdot} \bec{\nabla})\ln \rho$,
i.e., the anelastic approximation takes into account an inhomogeneous fluid density.

We study a long-term evolution of the particle number density in spatial scales $L_n$ which are much larger than the integral scale of turbulence $\ell_0$, and during the time scales $t_n$ which are much larger than the turbulent time scales $\tau_0$.
We use a mean-field approach in which all quantities are decomposed into the
mean and fluctuating parts, where the fluctuating parts have zero mean values, i.e., we use the Reynolds averaging.
In particular, the particle number density $n= \overline{n} + n'$,
where $\overline{n}=\langle n \rangle$ is the mean particle number density,
and $n'$ are particle number density fluctuations and $\langle n' \rangle=0$.
The angular brackets $\langle ... \rangle$ denote an ensemble averaging.
Averaging Eq.~(\ref{D1}) over an ensemble of a turbulent velocity field, 
we arrive at the mean-field equation for the particle number density:
\begin{eqnarray}
{\partial \overline{n} \over \partial t} + \bec{\nabla} {\bf \cdot} 
\left(\overline{\bm U} \, \overline{n}  + \langle {\bm u} \, n' \rangle \right) = D \, \Delta \overline{n} ,
\label{D5}
\end{eqnarray}
where ${\bm F} \equiv \langle {\bm u} \, n' \rangle$ is the turbulent flux of particles.
We consider for simplicity the case $\overline{\bm U}=0$.

To derive an expression for the turbulent flux of particles,
we obtain the equation for particle number density fluctuations $n'$,
by subtracting Eq.~(\ref{D5}) from Eq.~(\ref{D1}), which yields
\begin{eqnarray}
{\partial n' \over \partial t} = - \tilde{\cal Q} -({\bm u} {\bf \cdot} \bec{\nabla}) \overline{n} - \overline{n} (\bec{\nabla} {\bf \cdot} {\bm u}) + D \Delta n' ,
\label{D6}
\end{eqnarray}
where $\tilde{\cal Q} = \bec{\nabla} {\bf \cdot} \left(n' \, {\bm u} - \langle n' \, {\bm u} \rangle \right)$ is the nonlinear term. 
The source term for particle number density fluctuations, $-({\bm u} {\bf \cdot} \bec{\nabla}) \overline{n}$, 
results in a production of particle number density fluctuations by the tangling of the gradient $\bec{\nabla} \, \overline{n}$ of the mean particle number density by velocity fluctuations.
The other source term, $-\overline{n} (\bec{\nabla} {\bf \cdot} {\bm u})$ for particle number density fluctuations
can be rewritten as
$-\overline{n} (\bec{\nabla} {\bf \cdot} {\bm u}) = (\overline{n}/\overline{\rho} \,) ({\bm u} {\bf \cdot} \bec{\nabla}) \overline{\rho}$,
where we take into account the anelastic approximation,
$\bec{\nabla} {\bf \cdot} {\bm u} = - (1/\overline{\rho}) \, ({\bm u} {\bf \cdot} \bec{\nabla}) \overline{\rho}$,
which is also valid for the mean fluid density, $\overline{\rho}$.
This implies that this source term describes
a production of particle number density fluctuations by the tangling of the gradient  $\bec{\nabla} \, \overline{\rho}$
of the mean fluid density by velocity fluctuations.
We use the P\'{e}clet number ${\rm Pe} = |\tilde{\cal Q}| / |D \Delta n'|$ defined as the dimensionless ratio of the absolute values of the nonlinear term $|\tilde{\cal Q}|$ to the diffusion term $|D \Delta n'|$. The P\'{e}clet number can be estimated as
${\rm Pe} = \ell_0 \, u_{\rm rms}/D$. We consider the case of large P\'{e}clet and Reynolds number.
Since the nonlinear equation~(\ref{D6}) cannot be solved exactly for arbitrary P\'{e}clet numbers, 
we consider the case of large P\'{e}clet and Reynolds numbers, which corresponds to our laboratory experiments.

We apply the Fourier transform only in a ${\bm k}$ space but not in a $\omega$ space, because
in a fully developed Kolmogorov-like turbulence, the turbulent time is universally
related to spatial scales.
We take into account the nonlinear terms in equations for velocity
and particle number density fluctuations and apply the spectral $\tau$ approach
\citep{O70,PFL76}  (see also Ref.~\cite{RI21} for detail discussion).

For simplicity, we consider a one-way coupling by 
taking into account the effect of turbulence on the particle number density, and neglecting 
the feedback effect of the particle number density on the turbulent fluid flow.
The one-way coupling approximation is valid when the spatial density
of particles $n \, m_p$ is much smaller than the fluid density $\rho$, where $m_p$ is the particle mass.
First, we consider non-inertial particles, which means that the particles move with the fluid velocity, i.e., 
the particle number density is a passive scalar.

We use a multi-scale approach \cite{RS75}, i.e., we consider the one-point
second-order correlation function as:
\begin{eqnarray}
&& \left\langle u_{i}(t,{\bm x}) \, n'(t,{\bm x}) \right\rangle \equiv \lim_{{\bm x} \to {\bm y}}
\left\langle u_{i}(t,{\bm x}) \, n'(t,{\bm y}) \right\rangle
\nonumber\\
&& = \lim_{{\bm r} \to 0} \int F_{i}(t,{\bm k},{\bm R}) \exp({\rm i} \, {\bm k} {\bf \cdot} {\bm r}) \,d {\bm  k} =
\int F_{i}(t,{\bm k},{\bm R}) \,d {\bm  k},
\nonumber\\
\label{B19}
\end{eqnarray}
where $F_{i}(t,{\bm k},{\bm R}) =  \int F_{i}(t,{\bm k},{\bm K}) \exp({\rm i} \, {\bm K} {\bf \cdot} {\bm R}) \,d {\bm  K}$
and $F_{i}(t,{\bm k},{\bm K})=\left\langle u_i(t,{\bm k} + {\bm  K} / 2) \, n'(t,-{\bm k} + {\bm  K} / 2) \right\rangle $.
Here the mean fields depend on ``slow'' variables ${\bm R} = ({\bm x} +  {\bm y}) / 2$, while fluctuations
depend on ``fast'' variables ${\bm r} = {\bm x} - {\bm y}$,
which correspond to large-scale and small-scale spatial variables, respectively.
In the Fourier space, ${\bm k}= ({\bm k}_1 - {\bm k}_2) / 2$, corresponds to the small scales, and
${\bm K} = {\bm k}_1 + {\bm k}_2$ characterizes the large scales,
where we use the Fourier transform, $u_{i}(t,{\bm x}) =  \int u_{i}(t,{\bm k}_1) \exp({\rm i} \, {\bm k}_1 \cdot {\bm x}) \,d{\bm k}_1$.
For homogeneous turbulence, the correlation function, $F_{i}(t,{\bm k},{\bm R})$ is independent
of the large-scale variable ${\bm R}$, i.e., $F_{i}(t,{\bm k},{\bm R})=F_{i}(t,{\bm k})$.

To obtain expression for the particle turbulent flux, we use Eq.~(\ref{D6}) written in a Fourier space.
This allows us to derive equation for the correlation function
$F_j(t,{\bm k})$ in a Fourier space as
\begin{eqnarray}
{\partial F_j({\bm k}) \over \partial t} &=& - \left(\nabla_i \overline{n} - {\rm i} \, k_i \, \overline{n} \right) \, f_{ij}(-{\bm k})  + \hat{\cal M} F_j^{(III)}({\bm k}) ,
\nonumber\\
\label{F5}
\end{eqnarray}
where for the brevity of notation, hereafter we omit argument  $t$ in the correlation functions.
Here $\hat{\cal M} F_i^{(III)}({\bm k}) = \langle [\partial u_i(t, {\bm k}) / \partial t] \, n'(t, -{\bm k}) \rangle
- \langle u_i(t, {\bm k}) \, {\cal Q}(t, -{\bm k}) \rangle$
are the third-order moments appearing due to the nonlinear terms $\tilde{\cal Q}$
in Eq.~(\ref{D6}) and the nonlinear Navier-Stokes equation. Here
$f_{ij}({\bm k}) = \langle u_i(t, {\bm k}) \, u_j(t, -{\bm k}) \rangle$
and ${\cal Q}=\tilde{\cal Q}- D \Delta n' $.

We use the spectral $\tau$ approximation \cite{O70,PFL76}.
This approximation postulates that the deviations of the third-moment terms, $\hat{\cal M} F^{(III)}({\bm k})$, from the contributions to these terms afforded by the background turbulence, $\hat{\cal M} F^{(III,0)}({\bm k})$, can be expressed through  the similar deviations of the second moments, $F({\bm k}) - F^{(0)}({\bm k})$:
\begin{eqnarray}
&&\hat{\cal M} F^{(III)}({\bm k}) - \hat{\cal M} F^{(III,0)}({\bm
k}) = - {1 \over \tau_r(k)} \, \Big[F({\bm k})
\nonumber\\
&& \quad - F^{(0)}({\bm k})\Big],
\label{F10}
\end{eqnarray}
where $\tau_r(k)$ is the scale-dependent relaxation time, which can be identified with the correlation time $\tau(k)$ of the turbulent velocity field for large Reynolds and P\'{e}clet numbers.
The functions with the superscript $(0)$ correspond to the background turbulence with a zero turbulent particle flux and a zero level of
particle number density fluctuations. Consequently, Eq.~(\ref{F10}) reduces to $\hat{\cal M} F_i^{(III)}({\bm k}) = - F_i({\bm k}) / \tau(k)$. Validation of the $\tau$ approximation
for different situations has been performed in various numerical simulations
\cite{BS05,BRRK08,RKKB11,BRK12,HKRB12,EKLR17,RKB18}
(see also Ref.~\cite{RI21} for detail discussion of the ranges of applicability of this approach).

We assume that the characteristic time of variation of the second moment $F_i({\bm k})$ is substantially larger than the correlation time $\tau(k)$ for all turbulence scales.
This allows us to use a steady-staye solution of Eq.~(\ref{F5}).
Applying the spectral $\tau$ approximation and using the steady-state solution of Eq.~(\ref{F5}), we obtain the following formula for the turbulent flux of particles, $F_j({\bm k})$ as
\begin{eqnarray}
F_j({\bm k})= - \tau(k) \, \left(\nabla_i \overline{n} - {\rm i} \, k_i \, \overline{n} \right) \, f_{ij}^{(0)}(-{\bm k}) ,
\label{FF11}
\end{eqnarray}
where since we consider a one-way coupling, we replace the function $f_{ij}({\bm k})$ by $f_{ij}^{(0)}({\bm k})$
in Eq.~(\ref{FF11}).

We use the following model for the second moments of turbulent velocity field $f_{ij}^{(0)}({\bm k}) \equiv \langle u_i({\bm k}) \, u_j(-{\bm k}) \rangle^{(0)}$ of an isotropic and homogeneous background turbulence in anelastic approximation in a Fourier space:
\begin{eqnarray}
&& f_{ij}^{(0)}({\bm k})= {\left\langle {\bm u}^2 \right\rangle\, E(k) \over 8 \pi k^2} \biggl[\delta_{ij} -  k_{ij}
+ {{\rm i} \over k^2} \, \big(\lambda_i \, k_j - \lambda_j \, k_i\big)\biggr],
\nonumber\\
\label{F13}
\end{eqnarray}
where $k_{ij} =k_i \, k_j / k^2$, $\delta_{ij}$ is the Kronecker unit tensor,
${\bm \lambda} =- ({\bm \nabla} \overline{\rho})/ \overline{\rho}$,
the spectrum function of the turbulent kinetic energy density
is $E(k) = (2/3) \, k_0^{-1} \, (k / k_{0})^{-5/3}$
 [see Ref.~\cite{RI21} for detail derivation of Eq.~(\ref{F13})].
Here the wavenumber $k$ varies within the interval $k_0 \leq k \leq k_{\nu}$
corresponding to the inertial range of scales, the wave number $k_{0} = 1 / \ell_0$,
the length $\ell_0$ is the integral scale of turbulence, the wave number $k_{\nu}=\ell_{\nu}^{-1}$, where $\ell_{\nu} = \ell_0 {\rm Re}^{-3/4}$ is the Kolmogorov (viscous) scale
and the turbulent correlation time is given by $\tau(k) = 2 \, \tau_0 \, (k / k_{0})^{-2/3}$,
where $\tau_0$  is the characteristic turbulent time.
The functions $E(k)$ and $\tau(k)$ correspond to fully developed turbulence with the Kolmogorov scalings.

Substituting Eq.~(\ref{F13}) into Eq.~(\ref{B19}), we determine the turbulent flux of particles $F_i=\langle u_i \, n' \rangle$:
\begin{eqnarray}
F_i &=& - {\left\langle {\bm u}^2 \right\rangle \over 8 \pi} \, \int_{k_0}^{k_\nu} \, \tau(k) \, E(k)  \, dk \int_{0}^{2\pi} \, d\varphi \int_{0}^{\pi} \sin \vartheta \,d\vartheta
\nonumber\\
&&\times \left[\left(\delta_{ij} - k_{ij}  \right) \, \, \nabla_i \overline{n}
+ \left(\lambda_j - \lambda_i  \, k_{ij}  \right) \, \overline{n} \right] .
\label{F14}
\end{eqnarray}
For the integration over ${\bm k}$  in Eq.~(\ref{F14}), we use the integrals given by
$\int_{0}^{2\pi} \, d\varphi \int_{0}^{\pi} \sin \vartheta \,d\vartheta \,
k_{ij} = (4 \pi / 3) \, \delta_{ij}$ and
$\int_{k_0}^{k_\nu} \tau(k) \, E(k) \,dk = \tau_0$.
After integration over ${\bm k}$, we obtain
the particle turbulent flux $\left\langle n' \, {\bm u} \right\rangle$ as
\begin{eqnarray}
\left\langle n' \, {\bm u} \right\rangle = {\bm V}^{\rm eff} \, \overline{n} - D_T \, {\bm \nabla} \overline{n} ,
\label{F16}
\end{eqnarray}
where the turbulent diffusion coefficient is
\begin{eqnarray}
D_T = {1 \over 3}  \, \tau_0 \,  \left\langle {\bm u}^2 \right\rangle,
\label{F17}
\end{eqnarray}
and the effective pumping velocity is given by
\begin{eqnarray}
{\bm V}^{\rm eff} = - D_T {\bm \lambda} = D_T \, {{\bm \nabla} \overline{\rho} \over \overline{\rho}} .
\label{FFF17}
\end{eqnarray}
Equations~(\ref{F16})--(\ref{FFF17}) are in agreement with those obtained using dimensional analysis
(see Ref.~\cite{RI21} for detail discussions).
Note that the phenomenon of turbulent diffusion of particles has been predicted about 100 years ago in Ref.~\cite{T1922}.

To understand the mechanism related to the effective pumping velocity ${\bm V}^{\rm eff}$, we
use the equation of state for a perfect gas,
\begin{eqnarray}
P={k_B \over m_\mu} \, \rho \, T \equiv {R \over \mu} \, \rho \, T ,
\label{AA1a}
\end{eqnarray}
where $P$ is the fluid pressure, $k_B=R/N_A$ is the Boltzmann constant, $R$ is the gas constant,
$N_A$ is the Avogadro number, $\mu=m_\mu N_A$ is the molar mass,
and $m_\mu$ is the molecule mass.
We rewrite the equation of state for the mean fields assuming that $\overline{\rho} \, \overline{T} \gg \langle \rho' \, \theta \rangle$,
where $\rho' $ and $\theta$ are fluctuations of the fluid density and temperature, respectively, and 
$\overline{T}$ is the mean fluid temperature.
Thus, the equation of state for the mean fields reads:
\begin{eqnarray}
\overline{P}={k_B \over m_\mu} \, \overline{\rho} \, \overline{T} ,
\label{AA1b}
\end{eqnarray} 
where $\overline{P}$ is the mean pressure.

Using Eq.~(\ref{AA1b}), we express the gradient of the mean fluid density in terms of the gradients of the mean fluid pressure ${\bm \nabla} \overline{P}$ and mean fluid temperature ${\bm \nabla} \overline{T}$ as
\begin{eqnarray}
{{\bm \nabla} \, \overline{\rho} \over \overline{\rho}} = {{\bm \nabla} \overline{P} \over \overline{P}} - {{\bm \nabla} \overline{T} \over \overline{T}} .
\label{D13}
\end{eqnarray}
Substituting Eq.~(\ref{D13}) into Eq.~(\ref{FFF17}), we obtain the final expression for the effective
pumping velocity of non-inertial particles as
\begin{eqnarray}
{\bm V}^{\rm eff} = D_T \, \left({{\bm \nabla} \overline{P} \over \overline{P}} - {{\bm \nabla} \overline{T} \over \overline{T}}\right) .
\label{D14}
\end{eqnarray}

To understand different terms in Eq.~(\ref{D14}), we
compare the molecular and turbulent fluxes of particles
(or gaseous admixtures).
Equation for the number density of particles reads
\begin{eqnarray}
{\partial n \over \partial t} + \bec{\nabla} {\bf \cdot}(n \, {\bm U})
= - {\bm \nabla} \cdot {\bm F}_{\rm M},
\label{WW33}
\end{eqnarray}
where the molecular flux of particles ${\bm F}_{\rm M}$ is given by
\begin{eqnarray}
{\bm F}_{\rm M} = - D \left({\bm \nabla} n + k_{\rm t} { {\bm \nabla}
T \over T} + k_{\rm p} {{\bm \nabla} P \over P} \right),
\label{WW35}
\end{eqnarray}
which comprises three terms: molecular diffusion $(\propto {\bm
\nabla} n)$, molecular thermal diffusion for gases or thermophoresis
for particles $(\propto k_{\rm t} {\bm \nabla} T)$,
and molecular barodiffusion $(\propto k_{\rm p} {\bm \nabla} P$),
where $k_{\rm t}$ is the molecular thermal diffusion
ratio and $k_{\rm p}$ is the molecular barodiffusion
ratio.
Note that the phenomenon of  molecular thermal diffusion in gases has been predicted
long ago in Refs.~\cite{E1911,E1912,CH1912}.

In turbulent flows, the turbulent flux of particles
can be rewritten as
\begin{eqnarray}
{\bm F}_T \equiv \left\langle n' \,  {\bm u} \right\rangle = - D_T \left({\bm \nabla} \meann + \meann { {\bm \nabla} \meanT \over \meanT}  -  \meann {{\bm \nabla} \meanP \over \meanP} \right) ,
\label{WW34}
\end{eqnarray}
which is obtained by substitution of Eq.~(\ref{D14}) to Eq.~(\ref{F16}).
Comparing the molecular flux of particles~(\ref{WW35})
and the turbulent flux of particles~(\ref{WW34}), we can
interpret the new additional turbulent fluxes as fluxes caused by
the effects of turbulent thermal diffusion $[\propto k_T ({\bm
\nabla} \meanT)/\meanT]$ and turbulent barodiffusion $[\propto k_P ({\bf \nabla} \meanP)/\meanP]$,
where
\begin{eqnarray}
{\bm F}_T \equiv \left\langle n' \,  {\bm u} \right\rangle = - D_T \left({\bm \nabla} \meann + k_T \, { {\bm \nabla} \meanT \over \meanT}  +  k_P \, {{\bm \nabla} \meanP \over \meanP} \right) ,
\nonumber\\
\label{WWW34}
\end{eqnarray}
and $k_T = \meann$ is the turbulent thermal diffusion
ratio and $k_P = - \meann$ is the turbulent barodiffusion ratio.
These phenomena have been predicted in Refs.~\cite{EKR96,EKR97}.

For small inertial particles, the expression for the effective pumping velocity reads \citep{EKR98}
(see also Ref.~\cite{RI21} for detail derivation):
\begin{eqnarray}
{\bm V}^{\rm eff} =- \alpha \, D_T \,
{\bm \nabla} \ln \overline{T} ,
\label{G17}
\end{eqnarray}
where
\begin{eqnarray}
\alpha = 1 + 2 \, {m_p \over m_\mu} \, \left({\ln{\rm Re} \over {\rm Pe}} \right) \, {\overline{T}\over\overline{T}_\ast} = 1 + 2\, {V_g \, L_P \, \ln{\rm Re} \over u_0 \, \ell_0} ,
\nonumber\\
\label{G19}
\end{eqnarray}
where $\tau_p V_T^2 = \gamma \, (m_p / m_\mu) D = \gamma \, V_g \, L_P \, \overline{T}_\ast / \overline{T}$.
Here $\gamma=c_{\rm p}/c_{\rm v}$ is the ratio of specific heats,
$V_T=(\gamma \, k_B \overline{T}_\ast /m_\mu)^{1/2}$ is the thermal velocity,
$\overline{T}_\ast$ is the characteristic mean fluid temperature,
and $L_P = |\nabla_z \overline{P} / \overline{P}|^{\, -1}$ is the pressure height scale.
In derivation of Eq.~(\ref{G19}), we take into account that the Stokes time can be
written as $\tau_p = \overline{\rho} \, V_{\rm g} \, L_P / \overline{P}$
with ${\bm V}_{\rm g} = \tau_p \, {\bm g}$  being the terminal fall velocity of particles,
where ${\bm g}$ is the acceleration caused by the gravity field.
For large P\'eclet numbers, ${\rm Pe}\gg 1$,
the turbulent thermal diffusion coefficient $\alpha=1$ for non-inertial particles,
while for inertial particles $\alpha$ depends on the particle mass, the Reynolds and P\'eclet numbers.

The non-diffusive turbulent flux of particles, $\overline{n} \,
{\bm V}^{\rm eff}$, toward the mean temperature
minimum is the main reason for the formation of
large-scale inhomogeneous distributions of
inertial particles in temperature-stratified
turbulence.
The steady-state solution of the equation
for the mean number density of inertial particles,
\begin{eqnarray}
{\partial \overline{n} \over \partial t} + \bec\nabla {\bf
\cdot} \, \big[\overline{n} \, ({\bm V}_{\rm g} + {\bm
V}^{\rm eff}) - (D+D_T) \, \bec\nabla \overline{n} \big] =0 ,
\label{G10}
\end{eqnarray}
satisfying the boundary condition with a zero total particle flux
at the boundary, is given by
\begin{eqnarray}
{\overline{n} \over \overline{n}_{\rm b}} = \left({\overline{T}\over\overline{T}_{\rm b}}\right)^{-{\alpha D_T \over D+D_T}} \, \exp \left[-\int_{z_{\rm b}}^z \, {V_{\rm g}
\over D+D_T} \,d z' \right],
\label{G18}
\end{eqnarray}
where the subscripts $({\rm b})$ represent the values of the
mean temperature and the mean particle number density at the boundary $z=z_{\rm b}$.
Equation~(\ref{G18}) implies that
small inertial particles are accumulated below
the mean temperature minimum due to the gravity field.

The mechanism for turbulent thermal diffusion for inertial particles
is as following.
Particles inside the turbulent eddies due to its inertia tend to be drift
out to the boundary regions between the eddies due to the centrifugal inertial force.
Indeed, for large P\'eclet numbers, molecular diffusion of
particles in equation for the number density of inertial particles,
\begin{eqnarray}
{\partial n \over \partial t} + \bec{\nabla} {\bf \cdot} \left(n \, {\bm u}^{(p)} - D
\bec{\nabla} n \right) = 0 ,
\label{DDD2}
\end{eqnarray}
can be neglected, so that
\begin{eqnarray}
\bec\nabla {\bf \cdot} \, {\bm u}^{(p)} \approx - n^{-1} \, \left[{\partial n \over \partial t} + ({\bm u}^{(p)} \cdot {\bm \nabla}) n\right] \equiv  - n^{-1} \, {{\rm d}n \over {\rm d}t} ,
\nonumber\\
\label{ZZ30}
\end{eqnarray}
where ${\bm u}^{\rm(p)}$ is the particle velocity.
On the other hand, for inertial particles, $\bec\nabla {\bf \cdot} \, {\bm u}^{(p)} =
\bec\nabla {\bf \cdot} \, {\bm u} + (\tau_p / \overline{\rho})  \,\bec\nabla^2 p$.
Indeed, the solution of the equation of motion for inertial particles,
\begin{eqnarray}
{{\rm d} {\bm u}^{\rm(p)} \over {\rm d} t} =- {{\bm u}^{\rm(p)}-{\bm u} \over \tau_p} + {\bm g},
\label{G1}
\end{eqnarray}
 for $\rho_{\rm p} \gg \overline{\rho}$ and small Stokes time, reads  \cite{M87}: ${\bm u}^{\rm(p)} = {\bm u} - \tau_p \, ({\rm d} {\bm u} / {\rm d} t) + \tau_p{\bm g} + {\rm O}(\tau_p^2) \approx
{\bm u} + \tau_p \, ({\bm \nabla} p/ \overline{\rho}) + \tau_p{\bm g} + {\rm O}(\tau_p^2)$.
Here $\rho_{\rm p}$ is the material density of particles.
This yields the equation for $\bec\nabla {\bf \cdot} \, {\bm u}^{(p)}$.
Therefore, in regions with maximum fluid pressure fluctuations (where
$\bec\nabla^2 p < 0)$, there is accumulation of
inertial particles, i.e.,  ${\rm d} n' / {\rm d}t
\propto - \overline{n} \, (\tau_p /\overline{\rho}) \,\bec\nabla^2 p
> 0$. These regions obey low vorticity and high
strain rate.
Similarly, there is an outflow of inertial
particles from regions with minimum fluid
pressure.

In a homogeneous and isotropic turbulence with a zero gradient
of the mean temperature, there is no a preferential direction.
This implies that in a homogeneous and isotropic turbulence 
there is no large-scale effect of particle accumulation,
and the pressure (temperature) of
the surrounding fluid is not correlated with the turbulent velocity field. The only
non-zero correlation is $\langle({\bm u} \cdot {\bm \nabla})p\rangle$,
which contributes to the flux of the turbulent
kinetic energy density.

In a temperature-stratified turbulence, 
the turbulent heat flux does not vanish, 
so that fluctuations of fluid temperature $\theta$
and velocity ${\bm u}$ are correlated, i.e., $\langle
\theta \, {\bm u} \rangle\not=\bm{0}$. Fluctuations of
temperature result in pressure fluctuations, which
cause fluctuations of the particle number density.
Increase of the pressure of the surrounding fluid
is accompanied by an accumulation of particles,
and the direction of the turbulent flux of particles
coincides with that of the turbulent heat flux.
The turbulent flux of particles is directed toward the
minimum of the mean temperature.
This causes the formation of large-scale
inhomogeneous structures in the spatial
distribution of inertial particles in the
vicinity of the mean temperature minimum.
In the next sections we will study this phenomenon in the experiments
with a forced convective turbulence.

\section{Experimental setup}
\label{sect3}

\begin{figure}
\centering
\includegraphics[width=7.5cm]{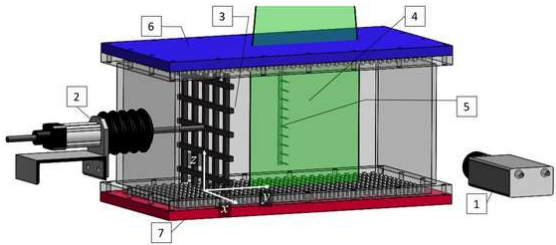}
\caption{\label{Fig1}
Experimental setup  with the forced convective turbulence: (1) digital CCD camera; (2) rod driven by the
speed-controlled motor; (3) oscillating grid;  (4) laser light sheet;
(5) temperature probe equipped with 12 E - thermocouples;
(6) heat exchanger at the top  cooled wall of the chamber;
(7) heat exchanger at the bottom heated wall of the chamber.
}
\end{figure}

In this section we describe the experimental set-up
and measurement technique.
We investigate turbulent
thermal diffusion of small solid particles in experiments with
a convective turbulence forced by one oscillating grid in the air flow.
We conduct experiments in rectangular transparent chamber
with dimensions $L_x \times L_y \times L_z$ with
$L_x=L_z=26$ cm and $L_y=53$ cm, where $Z$ is along the vertical
direction and $Y$ is perpendicular to the grid plain.
The oscillating grid with bars arranged in a square array is parallel to the side walls of the chamber, it is
positioned at a distance of two grid meshes from the left side wall of the chamber  (see Fig.~\ref{Fig1}).

Two aluminium heat exchangers with rectangular pins $3 \times 3 \times 15$
mm are attached to the bottom (heated) and top (cooled)
walls of the chamber,
which allow one to form a large vertical mean temperature gradient
up to 1.8~K/cm in the main fluid flow and about 7~K/cm close to the walls.
We measure the temperature field using a temperature probe equipped with 12
E - thermocouples.
The thermocouples with the diameter of 0.13 mm and
the sensitivity of $\approx 75 \, \mu$V/K are attached to
a vertical rod with a diameter 4 mm, and the mean distance between thermocouples
is about 21.6 mm (see for details Ref.~\cite{EKRL22}).
We measure the temperature field in many locations.
The data are recorded using the developed software based
on LabView 7.0, and the temperature maps are obtained
using Matlab~9.7.0.

We measure the velocity field with a
Particle Image  Velocimetry (PIV) system \cite{AD91,RWK07,W00},
consisting in a Nd-YAG laser (Continuum Surelite $2 \times
170$ mJ) and a progressive-scan 12 bit digital CCD
camera (with pixel size $6.45 \, \mu$m $\times \,
6.45 \, \mu$m and $1376 \times 1040$ pixels).
As a tracer for the PIV measurements,
we use an incense smoke with spherical solid particles
having the mean diameter of $0.7 \mu$m
and the material density $\rho_{\rm p}\approx 10^3 \rho$,
The particles are produced by high temperature sublimation of solid
incense grains (see for details Ref.~\cite{EKRL22}).

For instance, the velocity fields in our experiments have been measured  in a flow
domain $209.09 \times 155.43$ mm$^2$ with a spatial
resolution of $1376 \times 1024$ pixels, so that
a spatial resolution 151 $\mu$m /pixel have been achieved.
We analyse the velocity field in the probed region
with interrogation windows of $16
\times 16$ pixels.
Using the velocity measurements,
various  turbulence characteristics (e.g., the mean and the root mean square (r.m.s.)
velocities, two-point correlation functions and an integral scale of turbulence)
have been obtained in our experiments.
In particular, we determine the mean and r.m.s. velocities for
every point of a velocity map by averaging over 530 independent maps.
We obtain also the integral length scales of turbulence $\ell_y$ and $\ell_z$
in the horizontal $Y$ and the vertical $Z$ directions from the
two-point correlation functions of the velocity
field.

Next, we obtain the particle spatial distribution by the PIV system
using the effect of the Mie light scattering by particles \cite{guib01}.
To this end, we determine the mean intensity of scattered light
in $80 \times 64$ interrogation windows with the size
$16 \times 16$ pixels.
This allows us to find the vertical distribution of the intensity of
the scattered light in 80 vertical strips composed of
64 interrogation windows.
In particular, we take into account that
the light radiation energy flux scattered
by small particles is given by $E_s \propto E_0 \Psi(\pi d_{\rm p}/\lambda; a_0;n)$.
Here $\Psi$ is the scattering function, $d_{\rm p}$ is the particle diameter, $\lambda$ is the
wavelength, $a_0$ is the index of refraction.
The energy flux incident at the particle is given by
$E_0 \propto \pi d_{\rm p}^2 / 4$.
Note that when $\lambda > \pi d_{\rm p}$, the  scattering function $\Psi$ is determined
by the Rayleigh's law, $\Psi \propto d_{\rm p}^4$.
In opposite case for small $\lambda$, the scattering function $\Psi$ is independent of
the particle diameter and the wavelength.
In a general case, the scattering function $\Psi$ is determined by the Mie
equations \cite{BH83}.

Finally, we take into account that the light radiation energy flux scattered
by small particles is $ E_s \propto E_0 \, n \,
(\pi d_{\rm p}^2 / 4) $.
This implies that the scattered light energy flux
incident on the charge-coupled device (CCD) camera probe is proportional
to the particle number density $n$.
The ratio of the scattered radiation fluxes at two locations in the flow and at the
image measured with the CCD camera is equal to the ratio of the
particle number densities at these two locations.
For the normalization of the scattered light intensity $E^T$ obtained in
a temperature-stratified turbulence, we use
the distribution of the scattered light intensity E measured in the
isothermal case obtained under the same conditions.
Indeed, as follows from our measurements
applying different concentrations of the incense smoke,
the distribution of the scattered light intensity averaged over
a vertical coordinate is independent of the particle number
density in the isothermal flow.
Therefore, using this normalization, we can characterize the spatial distribution of particle
number density $n \propto E^T /E$ in the non-isothermal turbulence.

Note that the measurement technique and data processing procedure described in this section
are similar to those used by us  in various experiments with
turbulent convection \cite{BEKR09,EEKR11,SKRL22}
and stably stratified turbulence \cite{EEKR13,EKRL22}.
In addition, the similar measurement technique and data processing procedure
in the experiments have been performed previously by us to investigate
the phenomenon of turbulent thermal
diffusion in a homogeneous turbulence \cite{BEE04,EEKR04,EEKR06a,AEKR17}
as well as for study of small-scale particle clustering \cite{EKR10}.

\maketitle
\section{Experimental results}
\label{sect4}

\begin{figure}
\centering
\includegraphics[width=8.0cm]{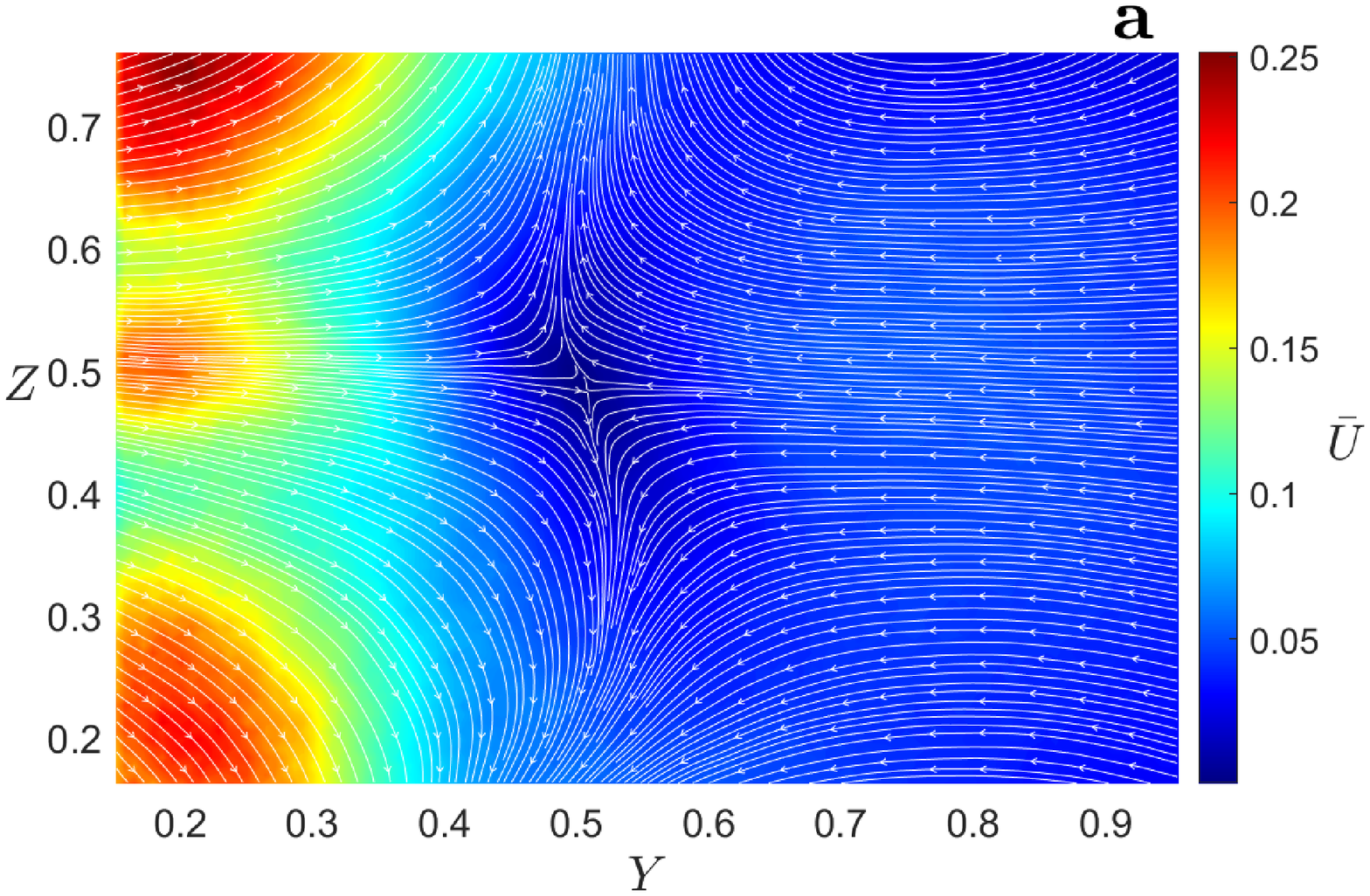}
\includegraphics[width=8.0cm]{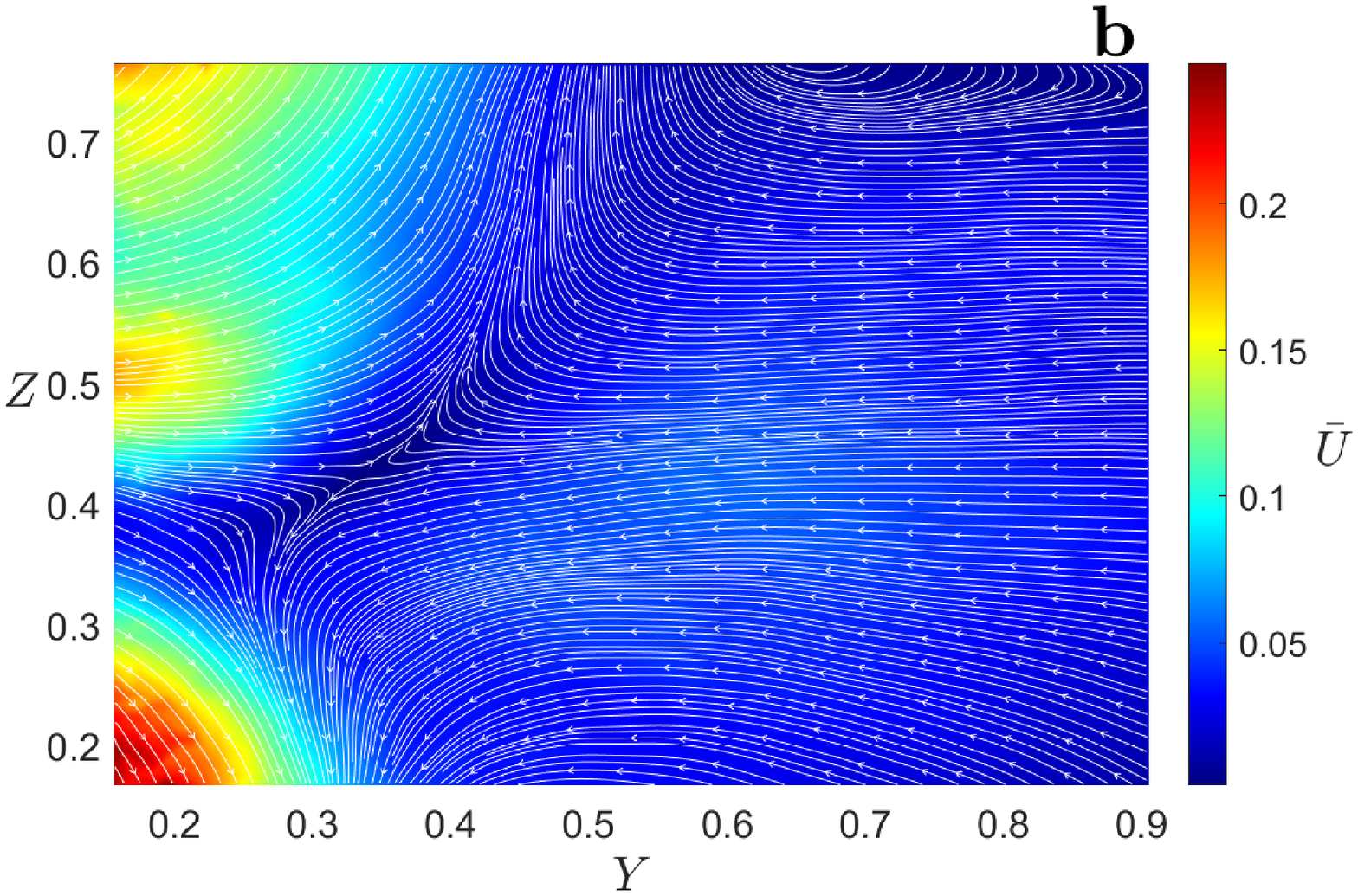}
\includegraphics[width=8.0cm]{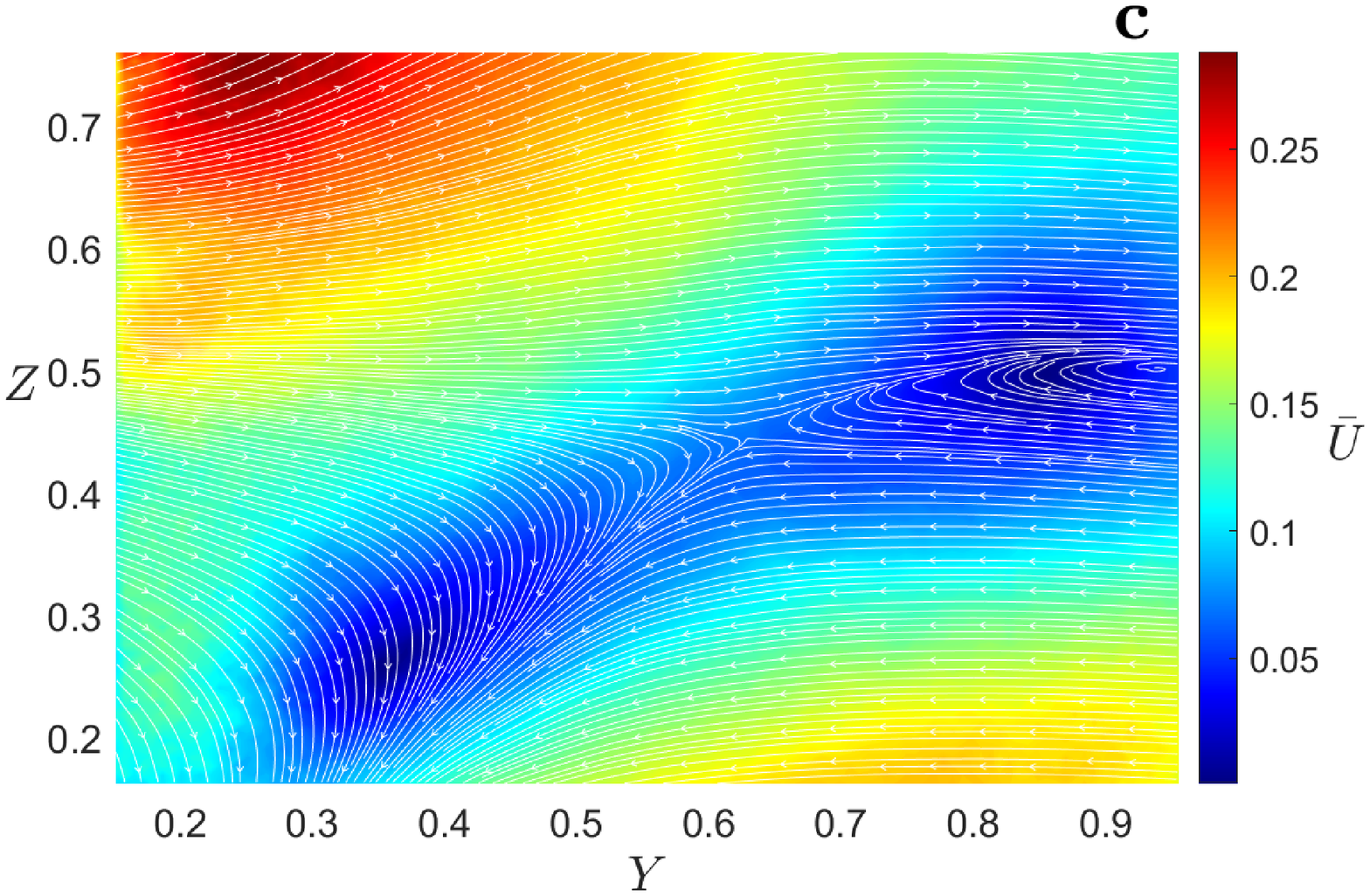}
\caption{\label{Fig3}
Mean velocity field  in the core flow for (a) isothermal turbulence;
(c) stably stratified turbulence; (b) forced convective turbulence. The velocity is measured
in m/s and coordinates $Y$ and $Z$ are normalized by $L_z=26$ cm.
}
\end{figure}

\begin{figure}
\centering
\includegraphics[width=7.0cm]{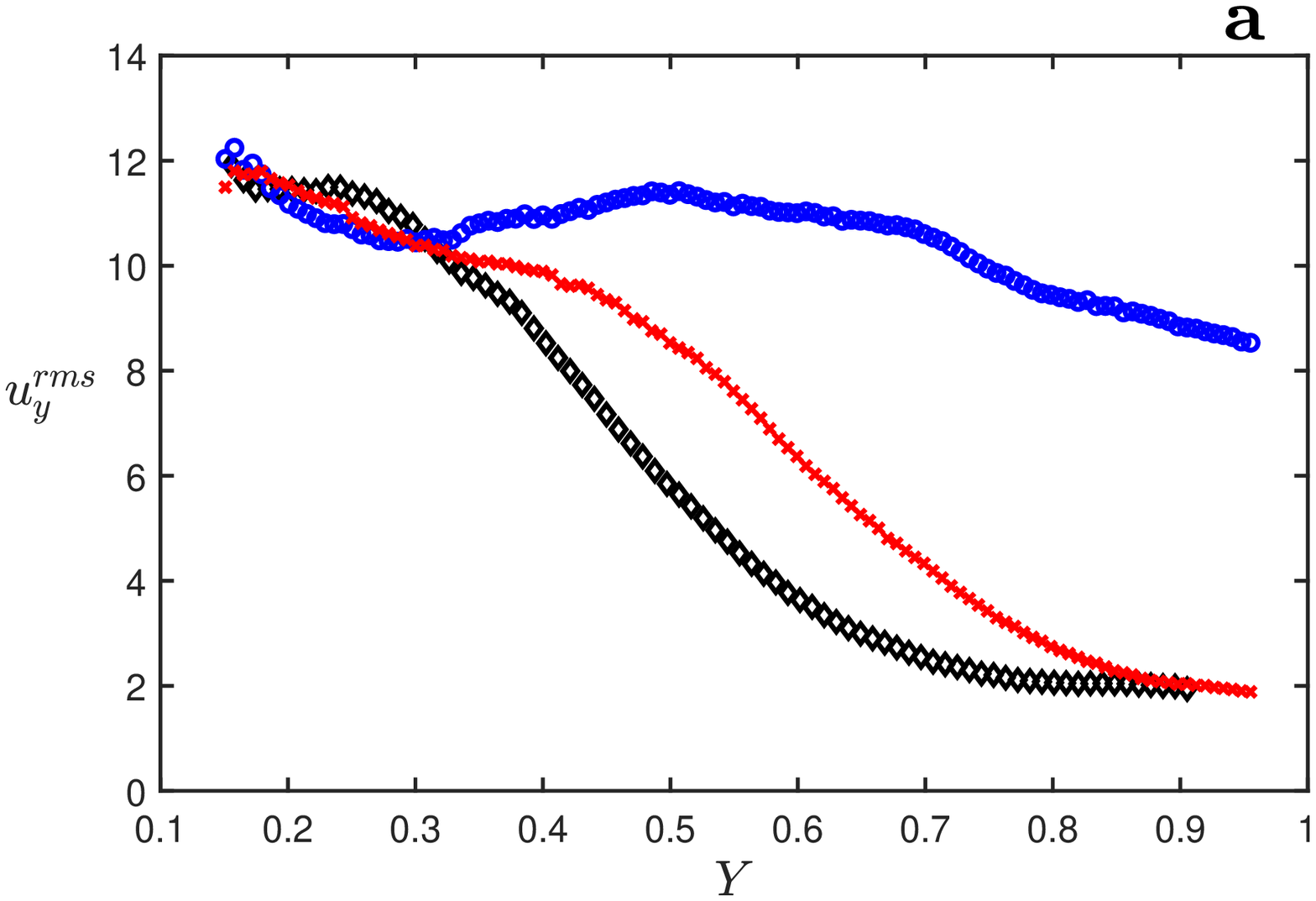}
\includegraphics[width=7.0cm]{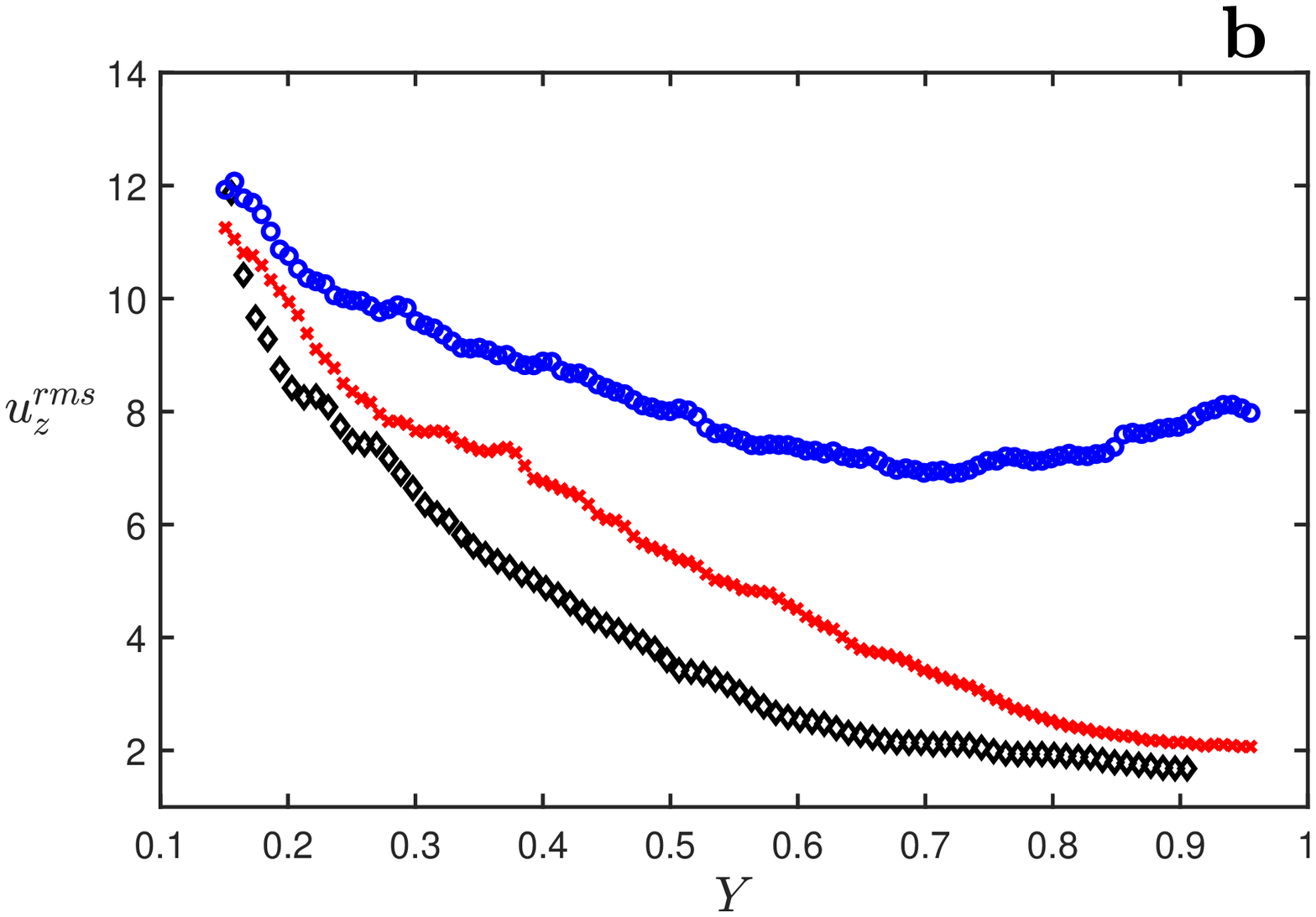}
\caption{\label{Fig4}
Horizontal  $u^{\rm (rms)}_y$ (upper panel) and vertical $u^{\rm (rms)}_z$ (lower panel)
components of the turbulent velocity versus normalized coordinate $Y$
averaged over the vertical coordinate $Z$
for isothermal turbulence (red);
stably stratified turbulence (black) and forced convective turbulence (blue).
The velocity is measured in cm/s and $Y$ is normalized by $L_z=26$ cm.
}
\end{figure}

In this section we discuss the obtained experimental results
in a forced convective turbulence with one oscillating grid in the air flow.
There are two sources of turbulence in a forced convective turbulence
with heated bottom wall of the chamber and cooling upper wall.
In particular, the turbulent kinetic energy is increased by buoyancy and
the grid oscillations.
In our experiments, the frequency $f$ of the grid oscillations is $f=10.5$ Hz,
which yields the maximum turbulence intensity in our experimental set-up.

Note that early laboratory experiments \cite{turn68,turn73,tho75,hop76,kit97,san98,med01}
which have been conducted in isothermal turbulence with one oscillating grid in a water flow
have demonstrated that the r.m.s. velocity behaves as $\sqrt{\langle {\bf u'}^2 \rangle} \propto f \, Y^{-1}$, 
while the integral turbulence length scale increases linearly with the distance $Y$ from a grid.
Therefore, the fluid Reynolds numbers as well as the turbulent diffusion
coefficient of particles are nearly independent of the distance $Y$ from the grid.
Our previous  \citep{EKRL22} and present studies in turbulence with one oscillating grid confirm these findings.

In the present study we conduct experiments in a forced convective turbulence with one oscillating grid
for the temperature difference $\Delta T =50$~K between the bottom and top walls of the chamber.
Using the PIV system we measure velocity field in the chamber 
for an isothermal and a forced convective turbulence, which allows us
to determine various turbulence characteristics.
In particular, we obtain the spatial distributions of the mean velocity in convective turbulence 
with large-scale circulations,
the vertical and horizontal profiles of the r.m.s turbulent velocity and  the integral turbulence length scales.
Since the oscillating grid is located near by the left wall of the chamber, and the amplitude of the grid oscillations is 6 cm,
we measure velocity field in the horizontal direction starting 20 cm away the left wall of the chamber.
We compare these  results with those obtained in our recent experiments  \citep{EKRL22} with
stably stratified turbulence produced by one oscillating grid in the air flow.

\begin{figure}
\centering
\includegraphics[width=8.0cm]{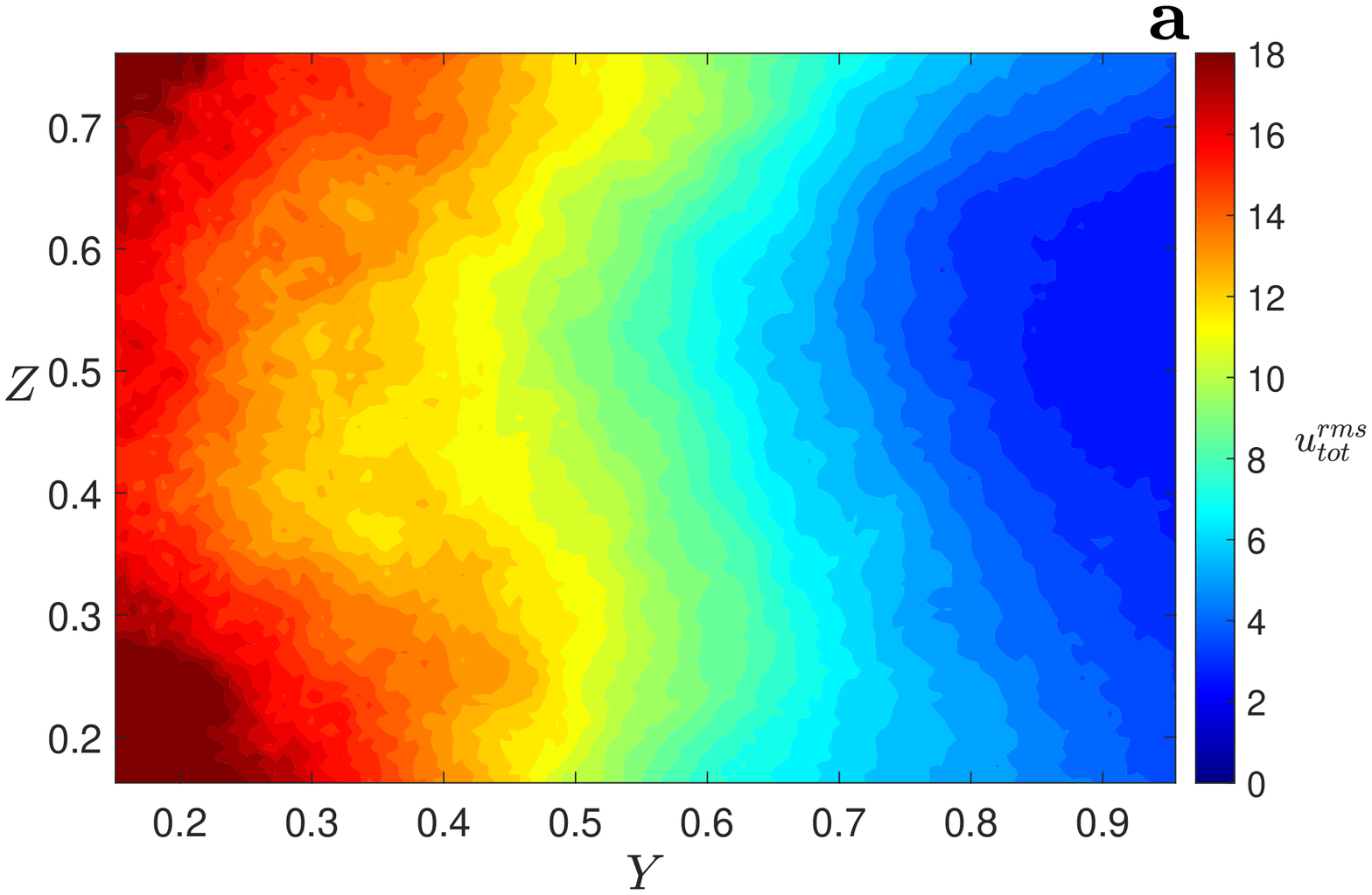}
\includegraphics[width=8.0cm]{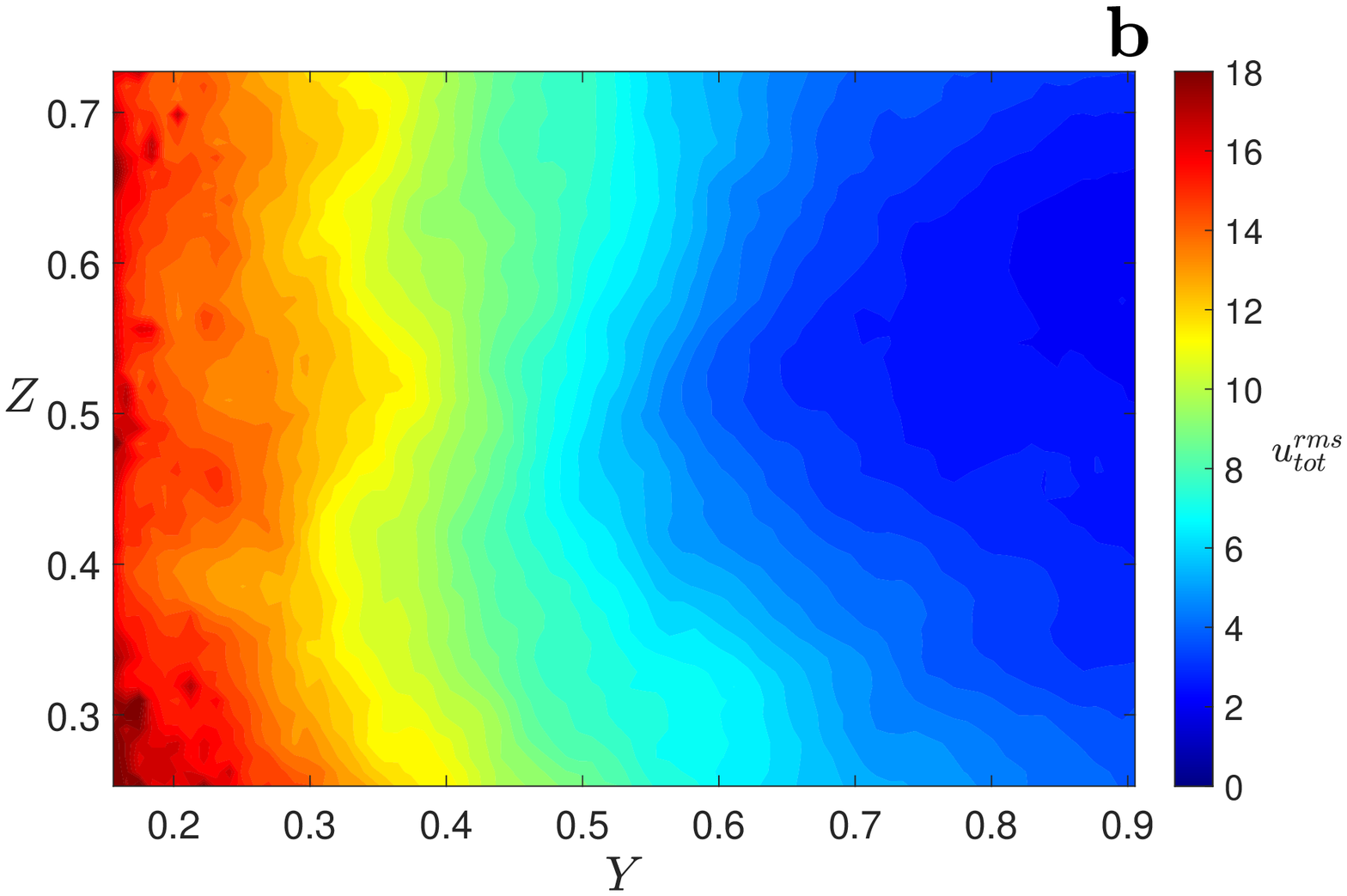}
\includegraphics[width=8.0cm]{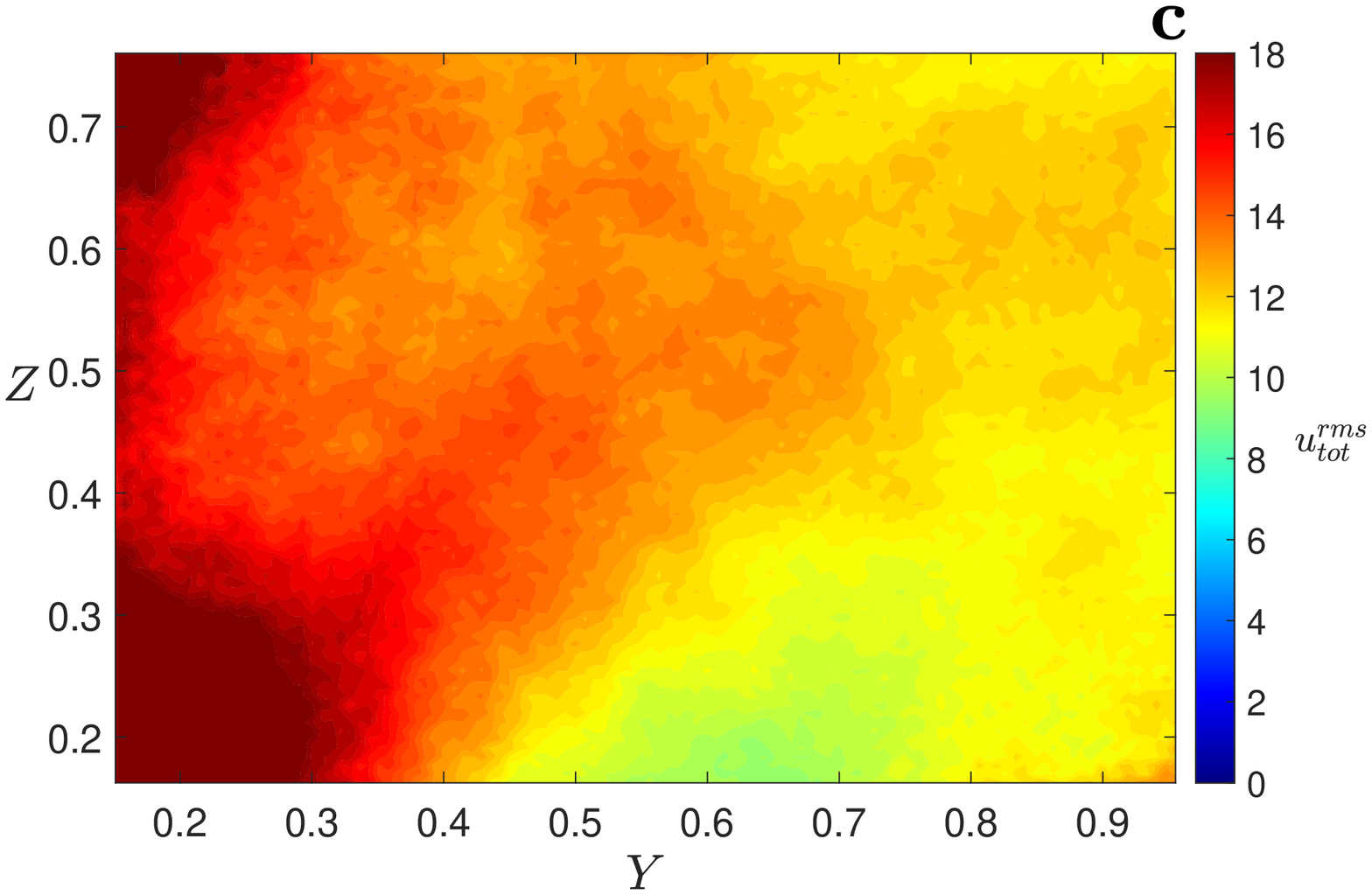}
\caption{\label{Fig5}
Distributions of the turbulent velocity $u^{\rm (rms)}_{\rm tot} = [\langle u_y^2 \rangle + \langle u_z^2 \rangle]^{1/2}$
for {\bf (a)} isothermal turbulence;
{\bf (b)} stably stratified turbulence and {\bf (c)} forced convective turbulence.
The velocity is measured in cm/s and coordinates are normalized by $L_z=26$ cm.
}
\end{figure}

Figure~\ref{Fig3} with the mean velocity patterns $\overline{U}$  in the main fluid flow for isothermal,
stably stratified turbulence and convective turbulence,
demonstrates that the temperature stratification and additional forcing strongly affect the mean velocity distributions.
Contrary to our previous experiments with a forced convection with two oscillating grids \cite{EEKR11},
the large-scale circulations in the convective turbulence with one oscillating grid
are not destroyed at the frequency $10.5$ Hz of the grid oscillations,
but their structure is strongly deformed (see the bottom panel in Fig.~\ref{Fig3}).

\begin{figure}
\centering
\includegraphics[width=8.5cm]{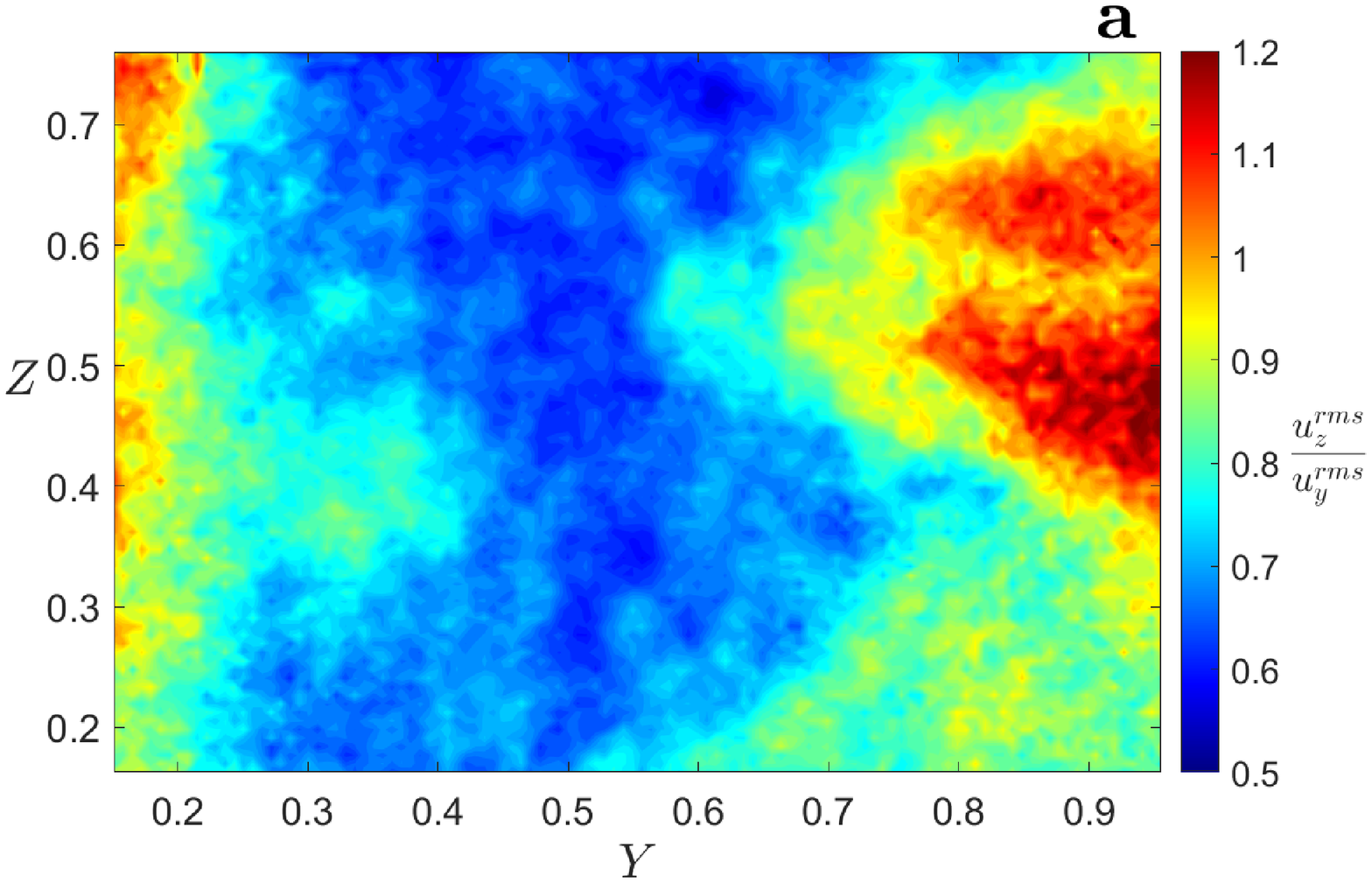}
\includegraphics[width=8.0cm]{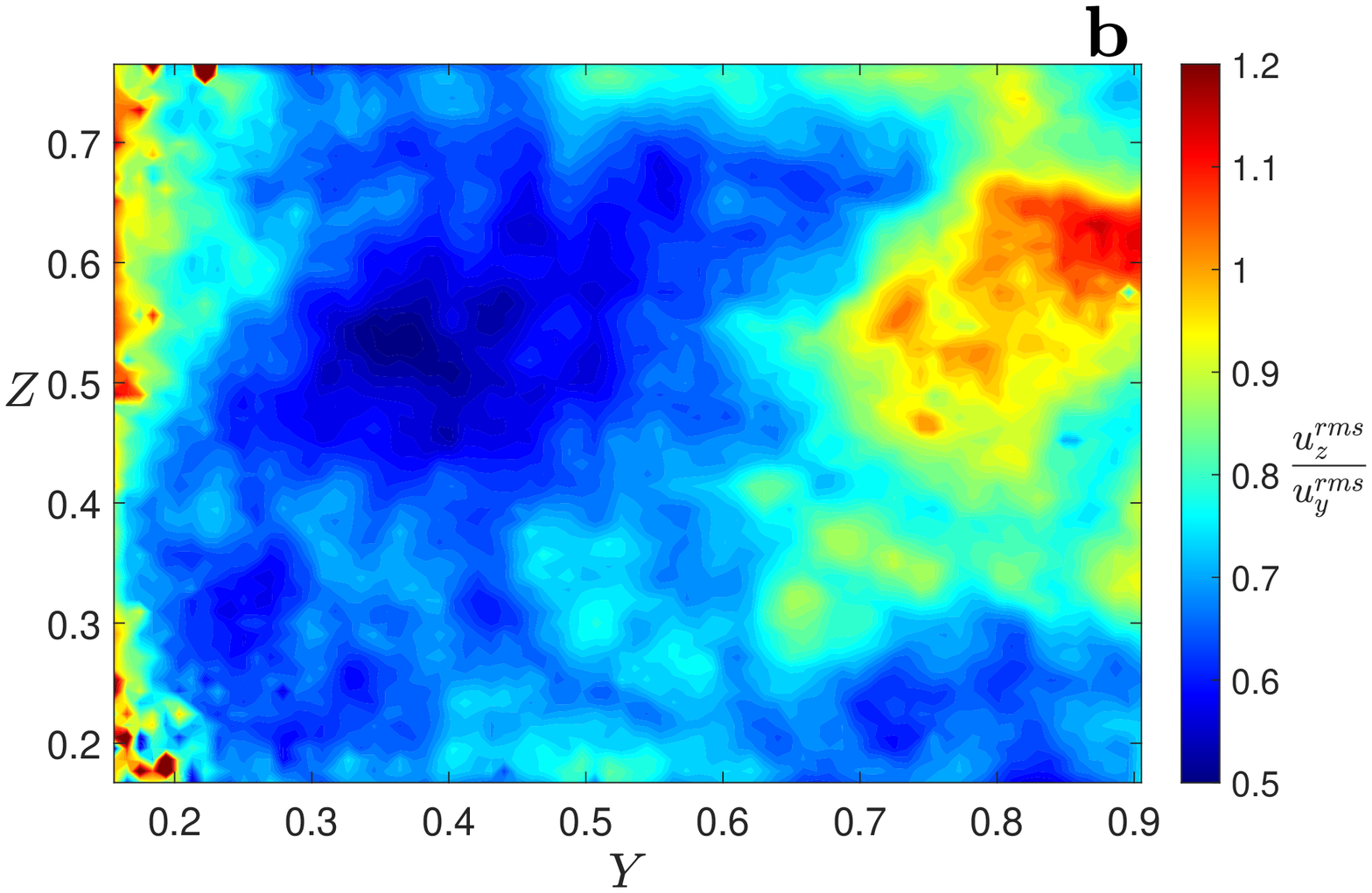}
\includegraphics[width=8.5cm]{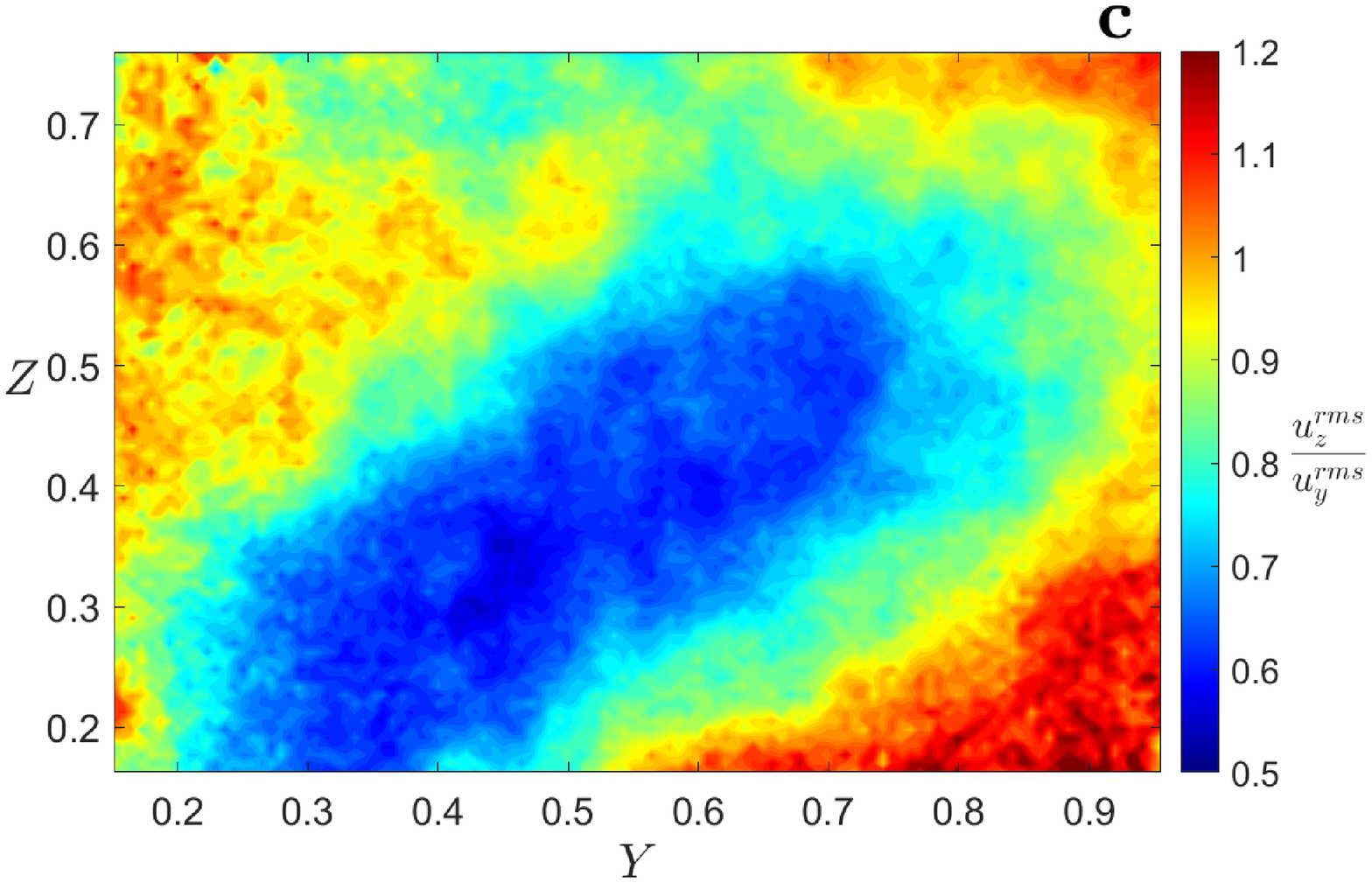}
\caption{\label{Fig6}
Distributions of the anisotropy parameter $u^{\rm (rms)}_z/u^{\rm (rms)}_y$ of the turbulent velocity
for {\bf (a)} isothermal turbulence;
{\bf (b)} stably stratified turbulence and {\bf (c)} forced convective turbulence.
The velocity is measured in cm/s and coordinates are normalized by $L_z=26$ cm.
}
\end{figure}

Similar effects of the temperature stratification and additional forcing are also seen in the horizontal profiles
of velocity fluctuations (see Fig.~\ref{Fig4}, where we plot the horizontal  $u^{\rm (rms)}_y$ 
and vertical $u^{\rm (rms)}_z$ components of turbulent velocities as the functions of $Y$
averaged over the vertical coordinate $Z$ for isothermal turbulence,
stably stratified turbulence and convective turbulence).
The turbulent velocities for convective turbulence are larger than for
isothermal turbulence, while the turbulent velocities for stably stratified turbulence are smaller than those for
isothermal and convective turbulence.
This is because the buoyancy increases the turbulent kinetic energy for convective turbulence
and decreases it for stably stratified turbulence.

\begin{figure}
\centering
\includegraphics[width=7.0cm]{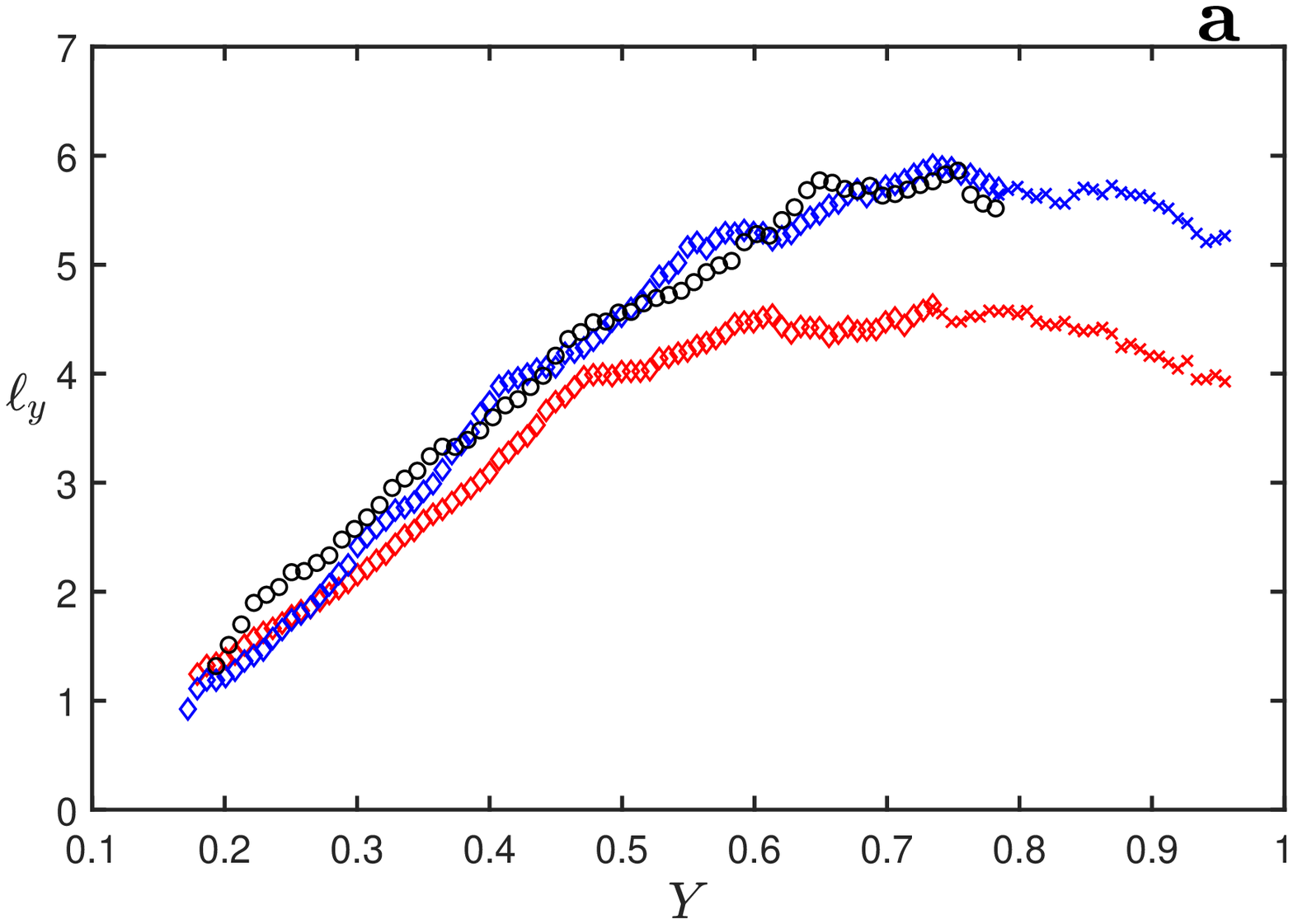}
\includegraphics[width=7.0cm]{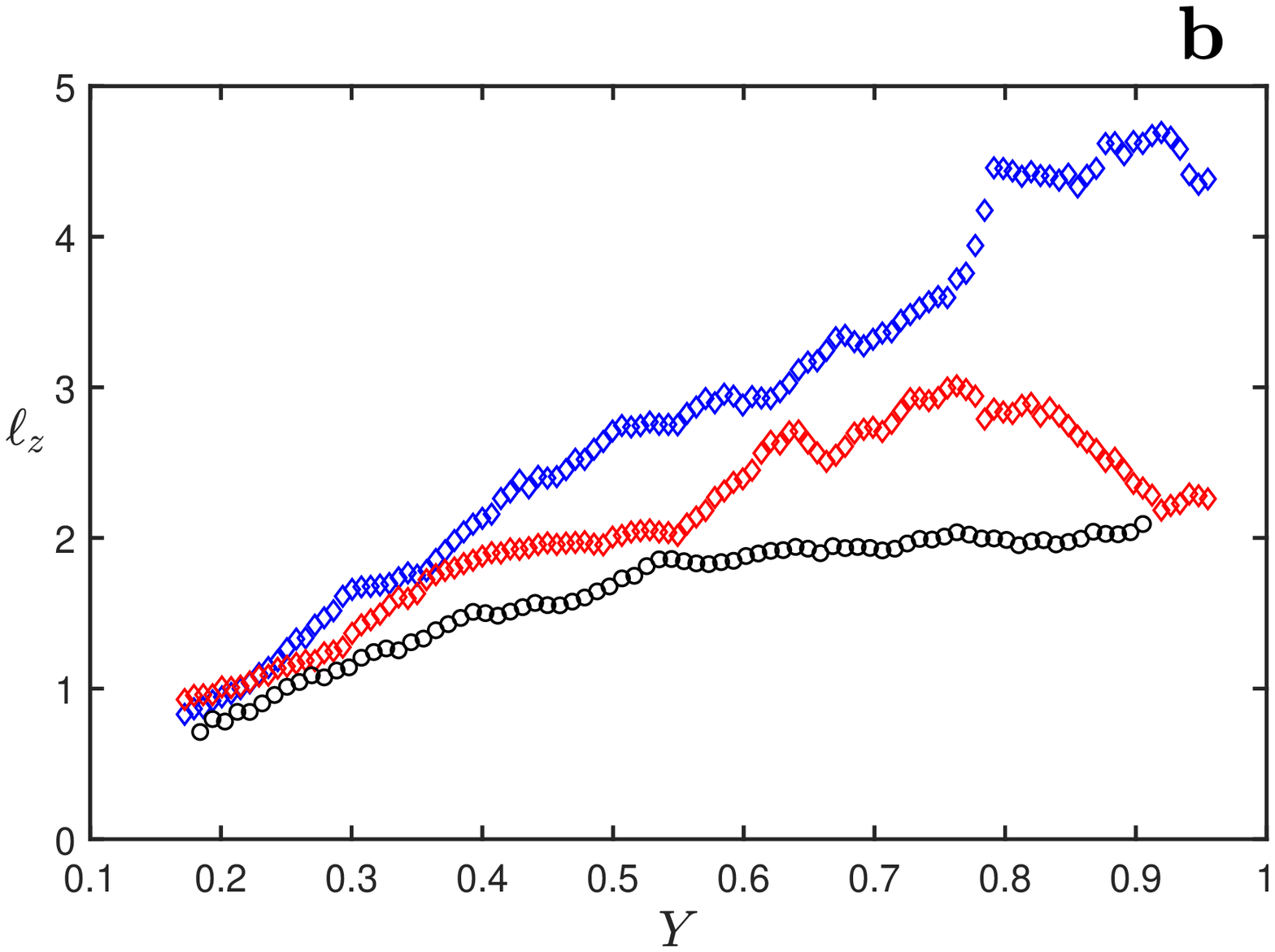}
\caption{\label{Fig7}
Horizontal  $\ell_y$ and vertical $\ell_z$ integral turbulent length scales versus normalized coordinate $Y$
averaged over $Z$
for isothermal turbulence (red);
stably stratified turbulence (black) and forced convective turbulence (blue).
The velocity is measured in cm/s and $Y$ is normalized by $L_z=26$ cm.
}
\end{figure}

\begin{figure}
\centering
\includegraphics[width=8.8cm]{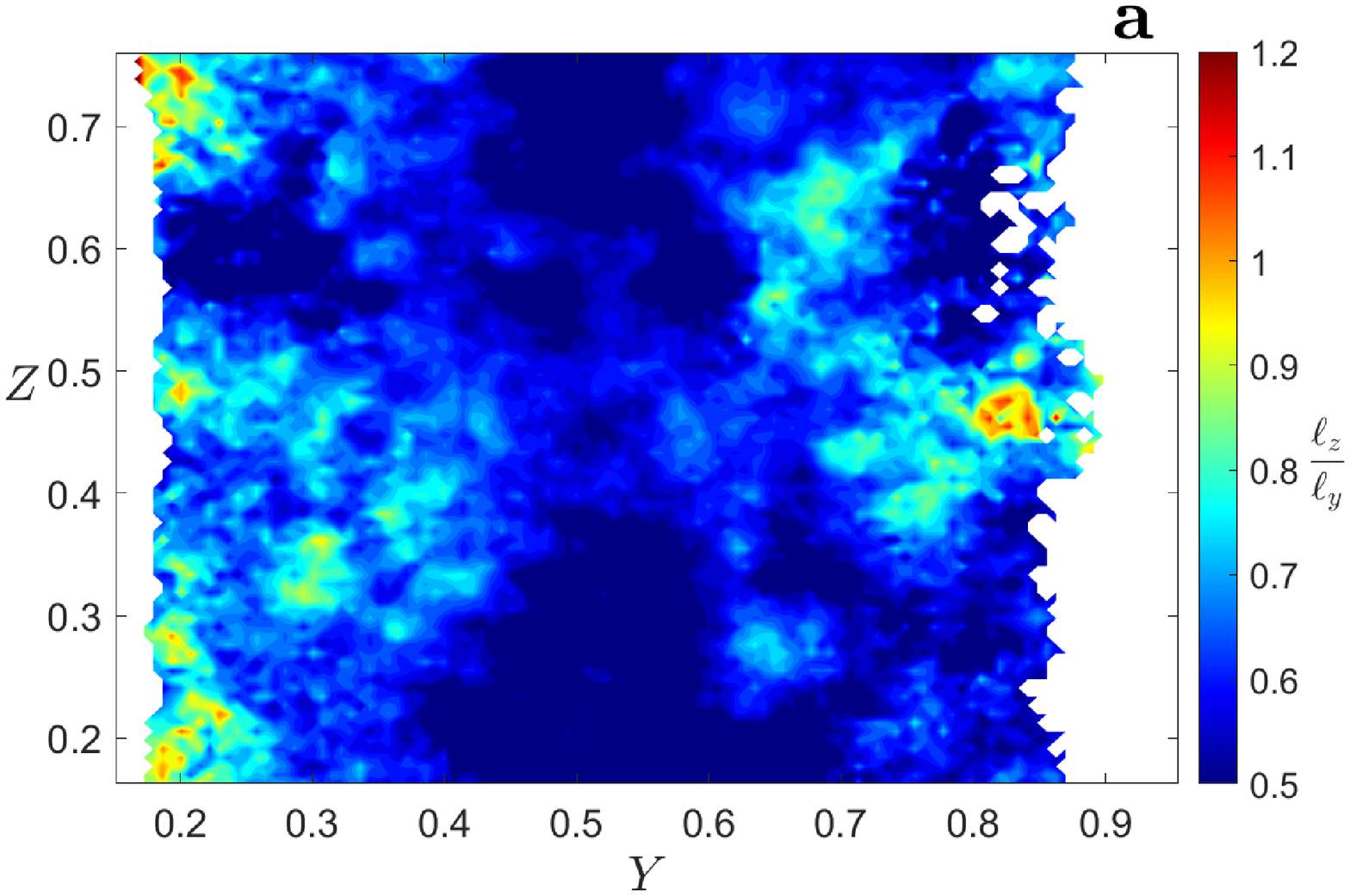}
\includegraphics[width=8.0cm]{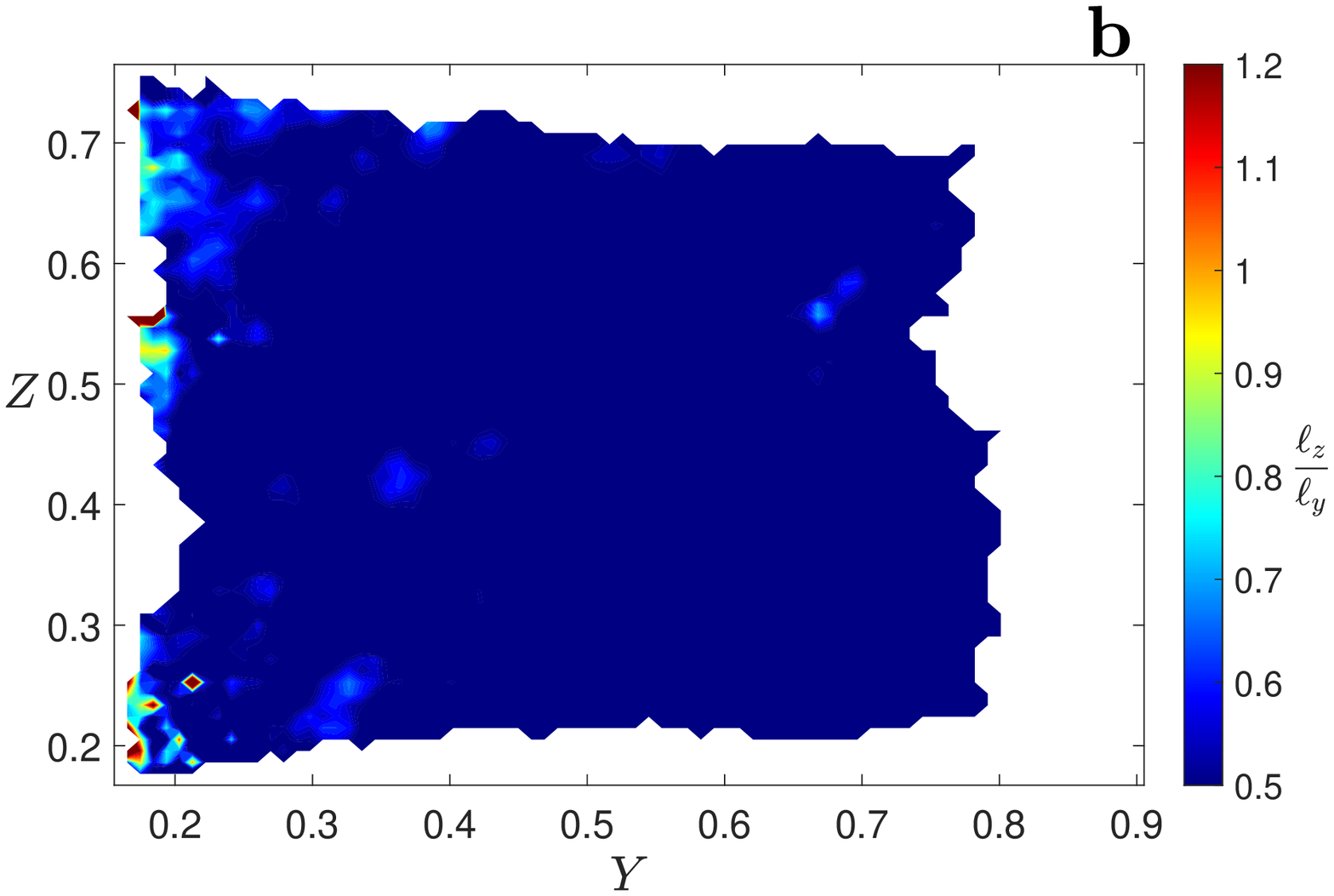}
\includegraphics[width=8.0cm]{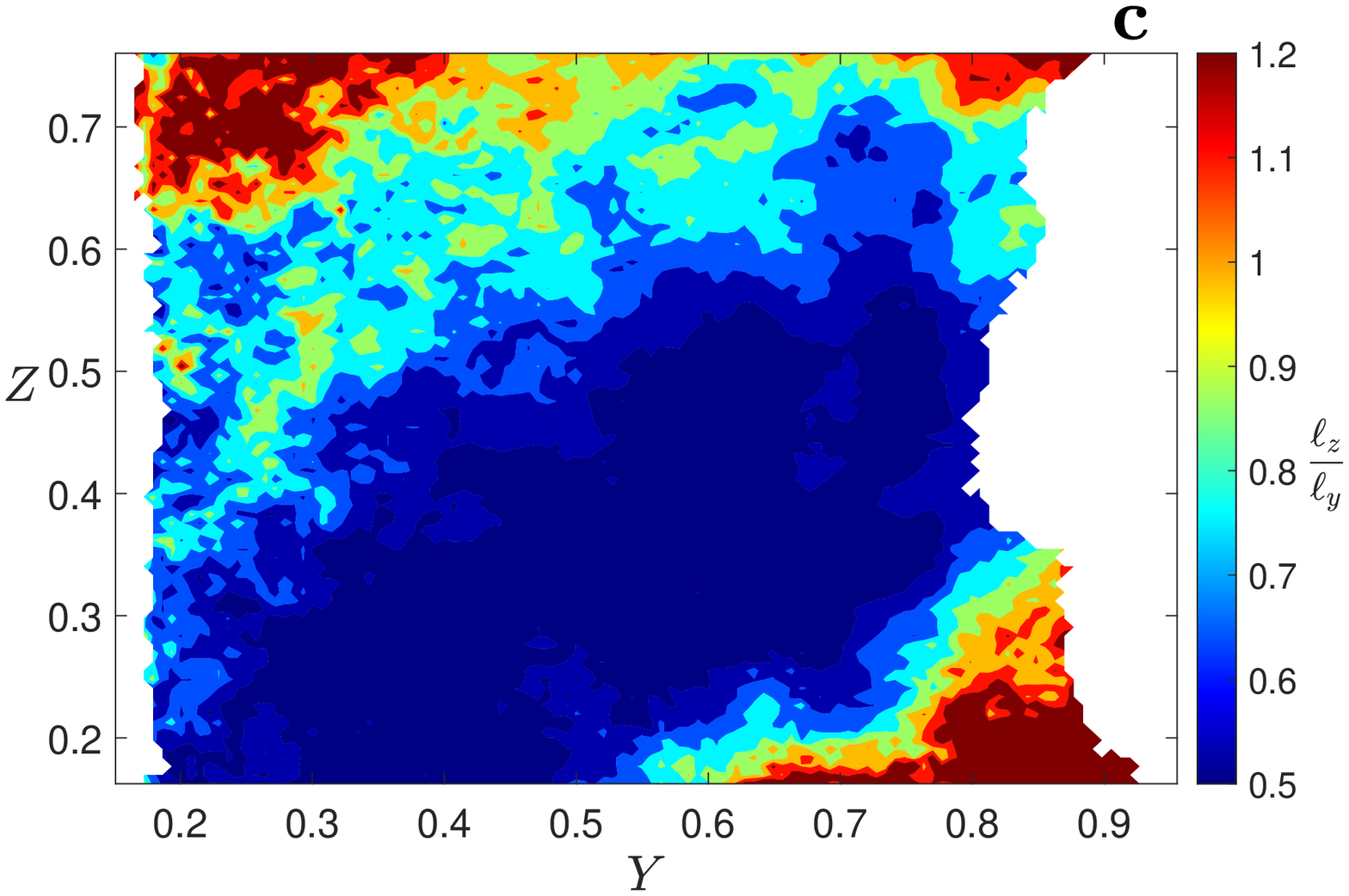}
\caption{\label{Fig8}
Distributions of the anisotropy parameter $\ell_z/\ell_y$ of the integral turbulent length scales
for {\bf (a)} isothermal turbulence;
{\bf (b)} stably stratified turbulence and {\bf (c)} forced convective turbulence.
The velocity is measured in cm/s and coordinates are normalized by $L_z=26$ cm.
}
\end{figure}

The oscillating grid strongly affects convective turbulence,
as can be seen in Figs.~\ref{Fig5} and ~\ref{Fig6}, where
we show the distributions of the turbulent velocity $u^{\rm (rms)}_{\rm tot} = [\langle u_y^2 \rangle + \langle u_z^2 \rangle]^{1/2}$
and  the anisotropy parameter $u^{\rm (rms)}_z/u^{\rm (rms)}_y$ for the turbulent velocity components
for isothermal , stably stratified and convective turbulence.
Figure~\ref{Fig6} demonstrates that
the anisotropy for isothermal and stably stratified turbulence
is more stronger than that for convective turbulence.
This is not surprising since the large-scale circulation enhances the mixing in the convective turbulence,
and it results in decrease of the turbulence anisotropy parameter $u^{\rm (rms)}_z/u^{\rm (rms)}_y$.
The same tendencies are also seen for the horizontal  $\ell_y$ and vertical $\ell_z$  integral turbulent length scales
shown in Fig.~\ref{Fig7}, as well as for the distributions of the anisotropy parameter
$\ell_z/\ell_y$ of the integral turbulent length scales (see Fig.~\ref{Fig8}).

To investigate the phenomenon of turbulent thermal diffusion
in a forced convective turbulence, we measure the spatial distributions 
of the mean temperature and the mean particle number density.
When small solid particles are injected into the chamber, their initial 
spatial distributions are nearly homogeneous.
Due to the effective pumping velocity 
caused by a combined effect of temperature-stratified turbulence and particle
inertia (described in terms of turbulent thermal diffusion) 
the final spatial distributions of the mean particle number density
is expected to be strongly inhomogeneous.

Sedimentation of particles can also result in a formation of inhomogeneous 
particle distributions near the bottom wall of the chamber.
However, this effect in our experiments is very weak,
because the terminal fall velocity for the micron-size particles is about $10^{-2}$ cm/s,
while the turbulent velocity in the experiments 
with the forced convective turbulence is much larger than 
the particle terminal fall velocity
(it is about $18$ cm/s near the grid and is more than $12$ cm/s far from the grid).
On the other hand, our estimates for the effective pumping velocity due to turbulent 
thermal diffusion shows that it is more than $3- 5$ cm/s near
the grid and is about $0.5$ cm/s far from the grid.
Note also that the Stokes time for for the micron-size particles is about $1.5 \times 10^{-6}$ s,
while in the forced convective turbulence the Kolmogorov time varies from $7 \times 10^{-3}$ s near the grid
up to $4 \times 10^{-3}$ s far from the grid.
Therefore, the turbulent and effective pumping velocities 
in our experiments are much larger than the terminal fall velocity for micron-size particles.

\begin{figure}
\centering
\includegraphics[width=8.0cm]{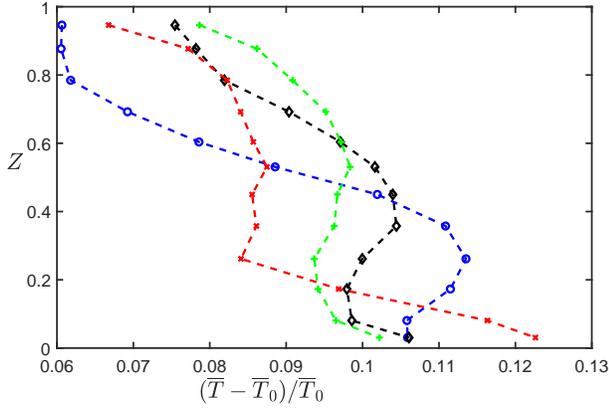}
\caption{\label{Fig9}
Vertical profiles of the relative normalized mean temperature $(\meanT - \meanT_0) / \meanT_0$
in the forced convective turbulence averaged over different horizontal regions:
$Y=4-10$ cm (blue, circles); $Y=11-16$ cm (black, diamond); $Y=17-25$ cm (green, crosses)
and $Y=24.5-28.5$ cm (red, slanting crosses), where $Z$ is normalized by $L_z=26$ cm.
}
\end{figure}

\begin{figure}
\centering
\includegraphics[width=8.0cm]{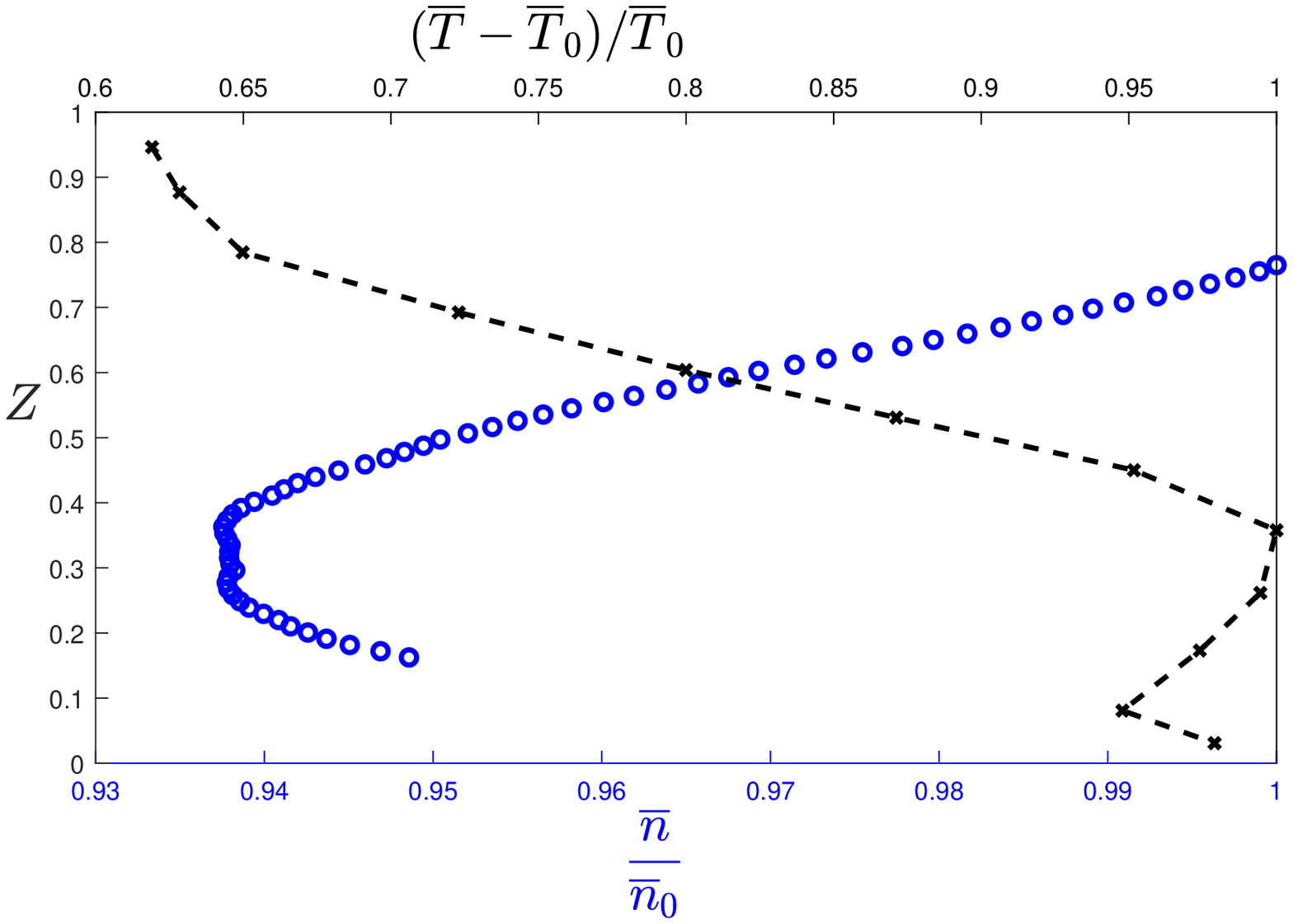}
\caption{\label{Fig10}
Vertical profiles of the relative normalized mean temperature
$(\meanT - \meanT_0) / \meanT_0$ (black, crosses)
and the normalized mean particle number density $\meanN(Y,Z) / \meanN_0$ (blue, circles)
in the forced convective turbulence averaged over horizontal region $Y=4-15$ cm (near the grid).
The coordinate $Z$ is normalized by $L_z=26$ cm.
}
\end{figure}

\begin{figure}
\centering
\includegraphics[width=8.0cm]{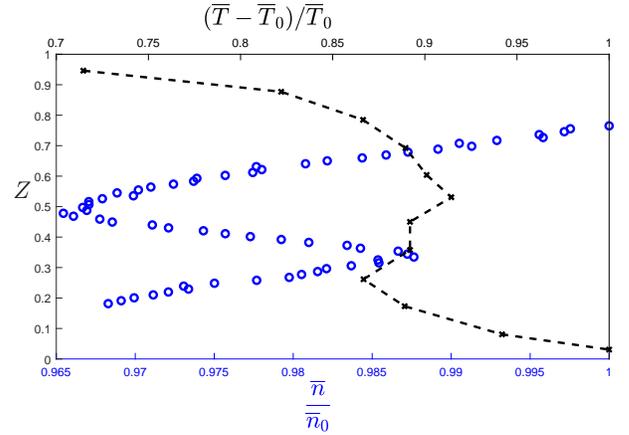}
\caption{\label{Fig11}
Vertical profiles of the relative normalized mean temperature
$(\meanT - \meanT_0) / \meanT_0$ (black, crosses)
and the normalized mean particle number density $\meanN(Y,Z) / \meanN_0$ (blue, circles)
in the forced convective turbulence averaged over horizontal region $Y=17-24$ cm (far from the grid).
The coordinate $Z$ is normalized by $L_z=26$ cm.
}
\end{figure}

Our experiments with a forced convective turbulence with large-scale circulations
show that the mean temperature is strongly nonuniform.
In particular, as follows from Fig.~\ref{Fig9} (where we plot vertical profiles of the relative normalized mean temperature
$(\meanT - \meanT_0) / \meanT_0$ averaged over different horizontal regions),
the normalized mean temperature near the grid increases with the height $Z$, reaches the maximum and
decreases nearly linearly with the height $Z$, where $\meanT_0$ is the reference mean temperature.
Far from the grid, the behavior of the mean temperature
is even more complicated, e.g., the normalized mean temperature
$(\meanT - \meanT_0) / \meanT_0$ decreases with the height $Z$, reaches the minimum and
increases with the height $Z$ reaching the maximum, and
finally it decreases nearly linearly with the height $Z$
(see Fig.~\ref{Fig9}).

To demonstrate the phenomenon of turbulent thermal diffusion
in the forced convective turbulence,  we show
in Figs.~\ref{Fig10}  and~\ref{Fig11} the vertical profiles of the relative normalized mean temperature
$(\meanT - \meanT_0) / \meanT_0$ (black crosses)
and the normalized mean particle number density $\meanN(Y,Z) / \meanN_0$ (blue circles)
near the grid (see  Figs.~\ref{Fig10}) and far from the grid (see  Figs.~\ref{Fig11}).
Due to the phenomenon of turbulent thermal diffusion,
the behaviour of the normalized mean particle number density
is opposite to the normalized mean temperature, i.e.,
the mean particle number density increases in the regions where the mean temperature decreases,
and the mean particle number density reaches the maximum at the minimum
of the mean temperature, and vise versa.
Therefore, Figs.~\ref{Fig10}  and~\ref{Fig11}
clearly demonstrate that particles are accumulated in the vicinity
of the minimum of the mean temperature even in very complicated
temperature field.

In the stably stratified turbulence, the behaviour of the mean temperature
and the mean particle number density
is more simple than for the forced convective turbulence \cite{EKRL22}.
In particular, the mean temperature
increases linearly with the height $Z$ in the flow for the stably
stratified turbulence, and the mean particle number density
decreases linearly with the height $Z$ due to the phenomenon of turbulent thermal diffusion.

\begin{figure}
\centering
\includegraphics[width=8.0cm]{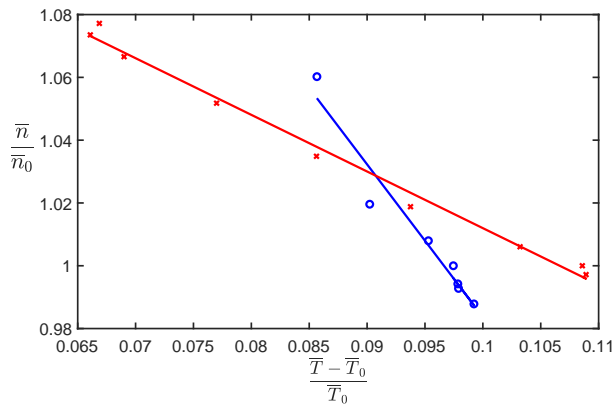}
\caption{\label{Fig12}
The normalized mean particle number density $\meanN / \meanN_0$ versus
the relative normalized mean temperature $(\meanT - \meanT_0) / \meanT_0$
averaged over different horizontal regions: $Y=4-15$ cm (blue, circles)
with $\alpha=4.86$
and $Y=16-24$ cm (red, crosses)
with $\alpha=1.74$
in the forced convective turbulence.
}
\end{figure}

To determine the effective turbulent thermal diffusion coefficient
$\alpha$ for particles in the forced inhomogeneous and anisotropic convective turbulence,
we show in Fig.~\ref{Fig12}
the normalized mean particle number density $\meanN / \meanN_0$ 
as the function of the relative normalized mean temperature $(\meanT - \meanT_0) / \meanT_0$,
where the slope of this dependence yields the coefficient $\alpha$.
In particular, we use a solution~(\ref{G18}) for Eq.~(\ref{G10}) for the mean particle number density $\meanN$
written in a steady-state, where we assume that $D_{\rm T} \gg D$ and neglect small terminal fall velocity.
Thus, we arrive at the following expression $\meanN / \meanN_0 = 1 - \alpha \, (\meanT - \meanT_0) / \meanT_0$,
which shows that the effective turbulent thermal diffusion coefficient
$\alpha$ for particles in the forced convective turbulence 
is $\alpha=4.86$ for particles accumulated in the regions $Y=4-15$ cm,
and $\alpha=1.74$ for particles accumulated in the regions $Y=16-24$ cm (see Fig.~\ref{Fig12}).
Now we take into account that turbulence far from the grid is less stronger than that near the grid.
Therefore, the effective turbulent thermal diffusion coefficient
$\alpha$ near the grid is larger than that far from the grid.
Therefore, this experimental study has demonstrated the effect of turbulent thermal diffusion
in an inhomogeneous and anisotropic forced convective turbulence.

\section{Conclusions}
\label{sect5}

In the present study, the effect of turbulent thermal diffusion of small solid particles,
resulting in formation of large-scale inhomogeneities in particle spatial distributions
in a temperature-stratified turbulence, has been investigated experimentally for micron-size particles
in an inhomogeneous convective turbulence forced by one oscillating grid  in the air flow.
The obtained experimental results have been compared
with the results of our previous experiments \cite{EKRL22,EEKR11} conducted in an inhomogeneous and anisotropic
stably stratified turbulence \cite{EKRL22} produced by a one oscillating grid
and in a forced convection with two oscillating grids in the air flow \cite{EEKR11}.
We have found that contrary to our previous experiments
with a forced convection with two oscillating grids \cite{EEKR11},
the large-scale circulations in the convective turbulence with a one oscillating grid
are not destroyed at the maximum frequency $10.5$ Hz of the grid oscillations,
but their structure is deformed (see Fig.~\ref{Fig3}).
The measured vertical turbulent velocities for convective turbulence are stronger than for
both, isothermal turbulence and stably stratified turbulence produced by a one oscillating grid,
since the buoyancy increases the turbulent kinetic energy for convective turbulence
and decreases it for stably stratified turbulence.
These effects are also observed in the measured vertical integral turbulent length scales
obtained from the two-point correlation functions for velocity fluctuations.

To study phenomenon of turbulent thermal diffusion, we measure spatial distributions of the mean
temperature and mean particle number density in many locations.
We have found that in the convective turbulence  near the grid, the mean temperature
increases with the height reaching the maximum and then it
decreases nearly linearly with the increase of the height.
On the other hand, far from the grid the behavior of the mean temperature in the convective turbulence
is more complicated. The mean fluid temperature decreases with the height
reaching the minimum, and for larger  heights it increases with the height reaching the maximum, and
finally it decreases nearly linearly with the height (see Fig.~\ref{Fig9}).

The behaviour of the mean particle number density
is opposite to the mean temperature.
In particular, the mean particle number density increases in the regions where the mean temperature decreases,
reaching the maximum near by the minimum of the mean temperature (see Figs.~\ref{Fig10}  and~\ref{Fig11}).
This implies that our experiments in convective and stably stratified turbulence with micron-size solid particles
have clearly demonstrated the existence of the phenomenon of turbulent thermal diffusion,
which causes particle accumulation in the vicinity of the minimum of the mean temperature
even in a complicated vertical profile of the mean  fluid  temperature.
We have determined the effective turbulent thermal diffusion coefficient
using the vertical profiles of the mean temperature and the mean particle number density.
We also have demonstrated that the obtained experimental results are in agreement
with the theoretical predictions.

\bigskip
\noindent
{\bf DATA AVAILABILITY}
\medskip

The data that support the findings of this study are available from the corresponding author
upon reasonable request.


\end{document}